\documentclass[12pt]{article}
\usepackage{amsthm,latexsym,multibox,amssymb}
\usepackage{graphicx,lscape,fancyhdr,array}
\usepackage[arrow,matrix,curve]{xy}\SilentMatrices
\def\xyma{\xymatrix@M.7em}
\def\xymas{\xymatrix@M.1em}
\newcommand{\be}{\begin{equation}}
\newcommand{\ee}{\end{equation}}
\newcommand{\ba}{\begin{eqnarray}}
\newcommand{\ea}{\end{eqnarray}}

\newtheorem{corollary}{Corollary}
\newtheorem{proposition}{Proposition}
\newtheorem{lemma}{Lemma}
\newcommand{\ft}{\footnotesize}
\newlength{\mylength}

\def\nn{\nonumber}

\def\a{\alpha}
\def\b{\beta}
\def\g{\gamma}

\def\d{\delta}

\def\k{\kappa}
\def\l{\lambda}

\def\m{\mu}
\def\n{\nu}
\def\r{\rho}
\def\s{\sigma}

\def\O{\Omega}
\def\g{\gamma}

\def\cs{{\mathcal S}}

\def\cl{{\cal L}}

\def\IR{\relax{\rm I\kern-.18em R}}
\def\ZZ{\relax{\hbox{\cmss Z\kern-.4em Z}}}

\csname @addtoreset\endcsname{equation}{section}
\newlength{\blength}
\renewcommand{\proof}[1]{\vspace{-.05cm}
\begin{list}{\bf Proof :}
{\listparindent=\parindent\parsep=0pt \labelwidth=-0.5cm
\labelsep=\parindent \addtolength{\labelsep}{-\blength}
\addtolength{\labelsep}{1.5cm}
\itemindent=-\blength
\addtolength{\itemindent}{\parindent} \leftmargin=1.0cm}
\item
#1~$\qedsymbol$\end{list} \vspace{.0cm}}
\begin{document}
\begin{titlepage}
\begin{flushright}

ULB--TH--02/23\\ hep-th/0208058\\

\end{flushright}
\vspace{1cm}

\begin{center} {\Large{\bf Tensor gauge fields in arbitrary representations of
$GL(D,\mathbb R)$ : duality $\&$ Poincar\'e lemma}}
\end{center}
\vspace{1cm}

\begin{center} {\large Xavier Bekaert and Nicolas Boulanger\footnote
{``Chercheur F.R.I.A.'', Belgium}
}
\end{center}
\vspace{.1cm}

\begin{center}{\sl
Facult\'e des Sciences, Universit\'e Libre de Bruxelles,\\
Campus Plaine C.P. 231, B--1050 Bruxelles, Belgium }\end{center}

\begin{center}
{xbekaert@ulb.ac.be ~~  nboulang@ulb.ac.be}
\end{center}

\vspace{1cm}
\begin{abstract}
Using a mathematical framework which provides a generalization of
the de Rham complex (well-designed for $p$-form gauge fields), we
have studied the gauge structure and duality properties of
theories for free gauge fields transforming in arbitrary
irreducible representations of $GL(D,\mathbb R)$. We have proven a
generalization of the Poincar\'e lemma which enables us to solve
the above-mentioned problems in a systematic and unified way.
\end{abstract}
\end{titlepage}
%
\section{Introduction}
%
The surge of interest in string field theories has refocused
attention on the old problem of formulating covariant field
theories of particles carrying arbitrary representation of the
Lorentz group. These fields appear as massive excitations of
string (for spin $S>2$). It is believed that in a particular phase
of M-theory, all such excitations become massless. The covariant
formulation of massless gauge fields in arbitrary representations
of the Lorentz group has been completed for $D=4$
\cite{Fronsdal:1978rb}. However, the generalization of this
formulation to arbitrary values of $D$ is a difficult problem
since the case $D=4$ is a very special one, as all the irreps of
the little group $SO(2)$ are totally symmetric. The covariant
formulation for totally antisymmetric representations in arbitrary
spacetime dimension has been easily obtained using differential
forms. For mixed symmetry type gauge fields, the problem was
partially solved in the late eighties \cite{Labastida:1989} (for
recent works, see for instance \cite{Burdik:2001hj}). A recent
approach \cite{Francia:2002} has shed new light on higher-spin
gauge fields, showing how it is possible to formulate the free
equations while foregoing the trace conditions of the
Fang-Fronsdal formalism. In this formulation, the higher-spin
gauge parameters are then not constrained to be irreducible under
$SO(D-1,1)$, it is sufficient for them to be irreducible under
$GL(D,\mathbb R)$.

Dualities are crucial in order to scrutinize non-perturbative
aspects of gauge field and string theories, it is therefore of
relevance to investigate the duality properties of arbitrary
tensor gauge fields. It is well known that the gravity field
equations in four-dimensional spacetime are formally invariant
under a duality rotation (for recent papers, see for example
\cite{Nieto:1999pn,Hull:2001}). As usual the Bianchi identities
get exchanged with field equations but, as for Yang-Mills
theories, this duality rotation does not appear to be a true
symmetry of gravity: the covariant derivative involves the gauge
field which is not inert under the duality transformation. A deep
analogy with the self-dual D$3$-brane that originates from the
compactified M$5$-brane is expected to occur when
 the six-dimensional (4,0) superconformal gravity theory is
compactified over a $2$-torus \cite{Hull:2000}. Thus, a
$SL(2,\mathbb Z)$-duality group for $D=4$ Einstein gravity would
be geometrically realized as the modular group of the torus. In
any case, linearized gravity does not present the problem
mentioned previously and duality is thus a true symmetry of this
theory. Dualizing a free symmetric gauge field in $D>4$ generates
new irreps of $GL(D,\mathbb R)$.

This paper provides a systematic treatment of the gauge structure
and duality properties of tensor gauge fields in arbitrary
representations of $GL(D,\mathbb R)$. Review and reformulation of
known results (\cite{Hull:2001,Labastida:1986} and the references
therein) are given in a systematic unified mathematical framework
and are presented together with new results and their proofs.

Section \ref{Lingravi} is a review of massless spin-two gauge
field theory and its dualisation. The obtained free dual gauge
fields are in representations of $GL(D,\mathbb R)$ corresponding
to Young diagrams with one row of two columns and all the other
rows of length one. Section \ref{N-c} ``N-complexes'' gathers
together the mathematical background needed for the following
sections. Based on the works
\cite{Olver:1983,Olver:1987,Dubois-Violette:1999,Dubois-Violette:2001},
it includes
 definitions and propositions together with a review of
linearized gravity gauge structure in the language of
$N$-complexes.
Section \ref{fieldequs} discusses linearized gravity field equations
and their duality properties in the introduced mathematical framework.
The section \ref{GenPl} presents our theorem,
 which
generalizes the standard Poincar\'e lemma. This theorem is then
used in section \ref{arbitraryYoung} to elucidate the gauge
structure and duality properties of tensor gauge fields in
arbitrary representations of $GL(D,\mathbb R)$. The proof of the
theorem, contained in the appendix, is iterative and simply
proceeds by successive applications of the standard Poincar\'e
lemma.
%
\section{Linearized gravity}\label{Lingravi}
%
\subsection{Pauli-Fierz action}
%
A free symmetric tensor gauge field $h_{\mu\nu}$ in $D$ dimensions
has the gauge symmetry
\be
\delta h_{\mu\nu}=2 \partial_{(\mu}\xi_{\nu)} \,.
\label{2gauge}
\ee
The linearized Riemann tensor for this field is
\be R_{\mu\nu\,\sigma\tau}\equiv\frac12(\partial_{\mu}
\partial_{\sigma}h_{\nu\tau}+\ldots)=-2\partial_{[\mu}h_{\nu][\sigma,\tau]}\,.
\label{Riemannlin} \ee It satisfies the property\be R_{\mu\nu\,
\sigma\tau}=R_{\sigma\tau\,\mu\nu}\label{symR} \ee together with
the first Bianchi identity \be R_{[\mu\nu\, \sigma]\tau} =0
\label{BianchiI} \ee and the second Bianchi identity \be
\partial_{[\rho}R_{\mu\nu]\, \sigma\tau}=0 \,.
\label{BianchiII}
\ee

It has been shown by Pauli and Fierz \cite{Fierz:1939} that there
is a unique, consistent action that describes a pure massless
spin-two field. This action is the Einstein action linearized
around a Minkowski background\footnote{Notice that the way back to
full gravity is quite constrained. It has been shown that there is
no local consistent coupling, with two or less derivatives of the
fields, that can mix various gravitons \cite{Boulanger:2000}. In
other words, there are no Yang-Mills-like spin-2 theories. The
only possible deformations are given by a sum of individual
Einstein-Hilbert actions. Therefore, in the case of one graviton,
\cite{Boulanger:2000} provides a strong proof of the uniqueness of
Einstein's theory.}\be S_{EH}[g_{\mu \nu}]=\frac{2}{\kappa^2}\int
d^D x \,\sqrt{-g}\,R_{full}\,,\quad g_{\m\n}=\eta_{\m\n}+\kappa
h_{\m\n}\,,\ee where $R_{full}$ is the scalar curvature for the
metric $g_{\m\n}$. The constant $\kappa$ has mechanical dimensions
$L^{D/2-1}$. The term of order $1/\kappa^2$ in the expansion of
$S_{EH}$ vanishes since the background is flat. The term of order
$1/\kappa$ is equal to zero because it is proportional to the
(sourceless) Einstein equations evaluated at the Minkowski metric.
The next order term in the expansion in $\kappa$ is the action for
a massless spin-2 field in $D$-dimensional spacetime \ba
S_{PF}[h_{\mu \nu}] &=& \int d^D x \left[
-\frac{1}{2}\left(\partial_{\mu}{h}_{\nu\rho}\right)\left(\partial^{\mu}
{h}^{\n\r}\right)
+\left(\partial_{\m}{h}^{\m}_{~\n}\right)\left(\partial_{\r}{h}^{\r\n}\right)
\right.\nonumber\\
&&\left.-\left(\partial_{\n}{h}^{\m}_{~\m}\right)\left(\partial_{\r}{h}^{\r\n}
\right)+\frac{1}{2}\left(\partial_{\m}{h}^{\n}_{~\n}\right)
\left(\partial^{\m}{h}^{\r}_{~\r}\right)\right] \, . \label{PF}
\ea Since we linearize around a flat background, spacetime indices
are raised and lowered with the flat Minkowskian metric $\eta_{\mu
\nu}$. For $D=2$ the Lagrangian is a total derivative so we will
assume $D\geq 3$. The (vacuum) equations of motion are the natural
free field equations \be R^\sigma{}_{\mu\, \sigma\nu}=0
\label{Einstein}\ee which are equivalent to the linearized
Einstein equations. Together with (\ref{Riemannlin}) the previous
equation implies that\be\partial^\mu R_{\mu\nu\,
\sigma\tau}=0\,.\label{d*R}\ee
%
\subsection{Minimal coupling}
%
The Euler-Lagrange variation of the Pauli-Fierz action is \be
\frac{\d S_{PF}}{\d h_{\m\n}}=R^{\s}_{\,\m\s\n}-\frac12\,
\eta_{\m\n}\,R^{\s\tau}_{\quad\s\tau}\,.\ee It can be shown that
the second Bianchi identity (\ref{BianchiII}) implies on-shell\be
\partial^\m R_{\m\s\n\r}+\partial_\r
R_{\n\m\s}^{\quad\,\,\m}-\partial_\n R_{\r\m\s}^{\quad\,\,\m}=0\,,
\ee and taking the trace again this leads to \be
\partial^\m R^{\s}_{\,\m\s\n}-\frac12\,
\partial_\n R^{\s\tau}_{\quad\s\tau}=0\,.\label{conservedR}\ee
From another perspective, the equations (\ref{conservedR}) can be
regarded as the Noether identities corresponding to the gauge
transformations (\ref{2gauge}).

Let us introduce a source $T_{\m\n}$ which couples minimally to
$h_{\m\n}$ through the term \be S_{minimal}=-\kappa\int d^D x \,
h^{\m\n}T_{\m\n}\,.\ee We add this term to the Pauli-Fierz action
(\ref{PF}), together with a kinetic term $S_K$ for the sources, to
obtain the action\be S=S_{PF}+S_K+S_{minimal}\,.\ee The field
equations for the symmetric gauge field $h_{\m\n}$ are the
linearized Einstein equations \be R^{\s}_{\,\m\s\n}-\frac12\,
\eta_{\m\n}\,R^{\s\tau}_{\quad\s\tau}= \kappa
T_{\m\n}\,.\label{Einsteinwsource}\ee Consistency with
(\ref{conservedR}) implies that the linearized energy-momentum
tensor is conserved $\partial^\m T_{\m\n}=0$.

The simplest example of a source is that of a free particle of
mass $m$ following a worldline $x^\m(s)$ with $s$ the proper time
along the worldline. The Polyakov action for the massive particle
reads \be S_{Polyakov}[x^\m(s)]= -m\int ds\,
g_{\m\n}\frac{dx}{ds}^\m \frac{dx}{ds}^\n \,.\ee It results as the
sum of the two actions \ba S_K&=&-m\int ds\,
\eta_{\m\n} \frac{dx}{ds}^\m \frac{dx}{ds}^\n \,,\label{kineticc}\\
S_{minimal}&=&-m\kappa \int ds \,h_{\m\n} \frac{dx}{ds}^\m
\frac{dx}{ds}^\n\,,\label{minimalc}\ea from which it can be
inferred that the (matter) source $T_{\m\n}$ for a massive
particle is equal to\be T^{\m\n}(x)=m\int ds
\,\d^{D}\left(x-x(s)\right)\, \frac{dx}{ds}^\m
\frac{dx}{ds}^\n\,.\ee This relationship is conserved if and only
if $\frac{d^2x}{ds^2}^\m=0$, which means that the test particle
follows a straight worldline. In general, when considering a free
massless spin-two theory coupled with matter, the latter has to be
constrained in order to be consistent with the conservation of the
linearized energy-momentum tensor\footnote{This should not be too
surprising since it is well known that the Einstein equations
simultaneously determine the gravity field \emph{and} the motion
of matter.}. At first sight, it is however inconsistent with the
natural expectation that matter reacts to the gravitational field.
Anyway, the constraint $\frac{d^2x}{ds^2}^\m=0$ is mathematically
inconsistent with the e.o.m. obtained from varying
(\ref{kineticc}) and (\ref{minimalc}) with respect to the
worldline $x^\m(s)$ which constrains the massive particle to
follow a geodesic for $g_{\m\n}$ (and \emph{not} a straight line).
In fact, for matter to respond to the gravitational field, it is
necessary to add a source $\kappa T^{self}_{\m\n}$ for the
gravitational field itself, in such a way that the sum
$T_{\m\n}+T^{self}_{\m\n}$ is conserved if the matter obeys its
own equation to first order in $\kappa$ and if the gravitational
field obeys (\ref{Einsteinwsource}). This gravitational
self-energy must come from a first order (in $\kappa$) deformation
of the Pauli-Fierz action. This modification was the starting
point of Feynman\footnote{In 1962, Feynman presented this
derivation in his sixth Caltech lecture on gravitation
\cite{Feynman}. One of the intriguing features of this viewpoint
is that the initial flat background is no longer observable in the
full theory. In the same vein, the fact that the self-interacting
theory has a geometric interpretation is ``\emph{not readily
explainable - it is just marvelous}", as Feynman expressed.} and
others in their derivation of the Einstein-Hilbert action by
consistent deformation of the Pauli-Fierz action with back reaction
\cite{Deser:1970}. At the end of the perturbative procedure, the
result obtained is that the free-falling particle must follow a
geodesic for consistency with the (full) Einstein equations.
%
\subsection{Duality in linearized gravity}\label{duallingravi}
%
Let us mention for further purpose that the equation
(\ref{d*R}) can be directly deduced from the equations
(\ref{BianchiI})-(\ref{BianchiII})-(\ref{Einstein}) for the
linearized Riemann tensor without using its explicit expression
(\ref{Riemannlin}). To simplify the proof and initiate a
discussion about duality properties, let us introduce the tensor
\be (*R)_{\mu_1\ldots\mu_{D-2}\,|\,\rho\sigma}=\frac12\,
\varepsilon_{\mu_1\,\ldots\,\mu_D}\,R^{\mu_{D-1}\mu_D}_{\quad\quad\quad
\rho\sigma}\label{*R}\,. \ee The linearized second Bianchi
identity and the Einstein equations can be written in terms of
this new tensor respectively as
\be\partial^\mu\left[(*R)_{\mu\,\ldots\nu\,|\,\rho\sigma}\right]=0
\label{EinsteinII} \ee and \be
(*R)_{[\mu\,\ldots\nu\,|\,\rho]\sigma}=0\,. \label{BianchiIV} \ee
Taking the divergence of (\ref{BianchiIV}) with respect to the
first index $\mu$, and applying (\ref{EinsteinII}), we obtain
\be\partial^\mu\left[(*R)_{\rho\,\ldots\nu\,|\,\mu\sigma}\right]=0\ee
which is equivalent to \be \partial^\mu R_{\a\b\,
\mu\sigma}=0\label{p*R}\ee as follows from the definition
(\ref{*R}). Using the symmetry property (\ref{symR}) of the
Riemann tensor we recover (\ref{d*R}).

In Corollary \ref{coro}, we will prove that the equations \be
R^\sigma{}_{\mu\, \sigma\nu}=0\,,\quad\partial^\mu
R_{\mu\nu\,\sigma\tau}=\partial^\sigma
R_{\mu\nu\,\sigma\tau}=0\label{d*R'}\ee are (locally) equivalent
to the following equation \cite{Hull:2001}
\be(*R)_{\mu_1\ldots\mu_{D-2}\,|\,\rho\sigma}=\partial_{[\mu_1}
\tilde{h}_{\mu_2\ldots\mu_{D-2}]\,|\,[\rho,\sigma]}\,,\label{dualcurva}\ee
which defines the tensor field
$\tilde{h}_{[\mu_1\ldots\mu_{D-3}]\,|\,\rho}$ called the
\emph{dual gauge field} of $h_{\m\n}$ and which is said to have
\emph{mixed symmetry} because it is neither (completely)
antisymmetric nor symmetric. In fact, it obeys the identity \be
\tilde{h}_{[\mu_1\ldots\mu_{D-3}\,|\,\rho]}\equiv 0\,. \ee
However, for $D=4$ the dual gauge field is a symmetric tensor
$\tilde{h}_{\mu\nu}$, which signals a possible duality symmetry.
The curvature dual (\ref{dualcurva}) remains unchanged by the
transformations
\be\delta\tilde{h}_{\mu_1\ldots\mu_{D-3}\,|\,\rho}=
\partial_{[\mu_1}S_{\mu_2\ldots\mu_{D-3}]\,|\,\rho}
+\partial_{\rho}A_{\mu_2\ldots\mu_{D-3}\mu_1}
+A_{\rho[\mu_2\ldots\mu_{D-3},\mu_1]}\label{gtransfoht}\ee where
complete antisymmetrization of the gauge parameter
$S_{[\mu_1\ldots\mu_{D-4}]\,|\,\mu_{D-3}}$ vanishes and the other
gauge parameter $A_{\mu_1\ldots\mu_{D-3}}$ is completely
antisymmetric.
%
\subsection{Mixed symmetry type gauge fields}\label{mixed}
%
Let us consider the general case of massless gauge fields
$M_{\m_1\m_2\ldots\m_n\,|\,\m_{n+1}}$ having the same symmetries
as the above-mentioned dual gauge field
$\tilde{h}_{\mu_1\ldots\mu_{D-3}\,|\,\rho}$. These can be
represented by the Young diagram 
\be 
\begin{picture}(30,40)(0,-20)
\multiframe(0,10)(10.5,0){1}(10,10){\ft$1$}
\multiframe(10.5,10)(10.5,0){1}(17,10){\ft$n\!\!+\!\!1$}
\multiframe(0,-0.5)(10.5,0){1}(10,10){\ft$2$}
\multiframe(0,-13.5)(10.5,0){1}(10,12.5){$ $}
\put(4,-12){$\vdots$}\put(40,-5){,}
\multiframe(0,-24)(10.5,0){1}(10,10){\ft$n$}
\end{picture}
\label{YS}
\ee
which implies that the field obeys the identity 
\be
M_{[\m_1\m_2\ldots\m_n\,|\,\m_{n+1}]}\equiv 0\,. 
\ee 
Such tensor
gauge fields were studied two decades ago by the authors of
\cite{Labastida:1986,Curtright:1985,Aulakh:1986} and appear in the
bosonic sector of some odd-dimensional CS supergravities
\cite{Troncoso:1998}. Here, $n$ is used to denote the number of
antisymmetric indices carried by the field
$M_{\m_1\m_2\ldots\m_n\,|\,\m_{n+1}}$, which is also the number of
boxes in the first column of the corresponding Young array. The
tensors $M_{\m_1\m_2\ldots\m_n\,|\,\m_{n+1}}$ have
$\frac{n(D+1)!}{(n+1)!(D-n)!}$ components in $D$ dimensions.

The action of the free theory is 
\be
S_0[M_{\m_1\m_2\ldots\m_n\,|\,\m_{n+1}}]=\int d^D x \,\cl
\label{startingpoint} 
\ee 
where the Lagrangian is\footnote{Notice
that, for $n=1$, the Lagrangian reproduces (\ref{PF})}
\cite{Aulakh:1986} 
\ba {\cal L}&=&M_{\m_1\ldots\m_n\,|\,\m_{n+1}}
\partial^2 M^{\m_1\ldots\m_n\,|\,\m_{n+1}}
 -2n M_{\m_1\ldots\m_n\,|\,\m_{n+1}}\partial^{\m_1}\partial^{\l}
M_{\l}^{~\m_2\ldots\m_n\,|\,\m_{n+1}}
\nonumber \\
&-&n M^{\m_1}_{~~\m_2\ldots\m_n\,|\,\m_1}\partial^{2}
M^{\m_1\ldots\m_n}_{~~~~~~~\m_1} +n(n-1)
M_{\m_1\m_2\ldots\m_n\,|\,\m}
\partial^{\m_1}\partial^{\n} M_{\n}^{~\m_2\ldots\m_{n-1}\m\,|\,\m_n}
\nonumber \\
&+&n(n-1) M_{\b\g\m_3\ldots\m_n}^{~~~~~~~~~\b}
\partial^{\g}\partial^{\m} M_{\n\m}^{~~\m_3\ldots\m_n\,|\,\n}
\nonumber \\
 &+&2n  M^{\m_1}_{~~\m_2\ldots\m_n\,|\,\m_1}\partial^{\m}\partial_{\n}
M_{\m}^{~\m_2\ldots\m_n\,|\,\n}\,.
\label{Lagrangian}
\ea
The field equations derived from (\ref{Lagrangian}) are equivalent to
\be
\eta^{\m_1\n_1} K_{\m_1\m_2\ldots\m_{n+1}\,|\,\n_1\n_2}=0\,
\ee
where 
\be 
K_{\m_1\m_2\ldots\m_{n+1}\,|\,\n_1\n_2}\equiv
\partial_{[\m_1}M_{\m_2\ldots\m_{n+1}]\,|\,\,[\n_1\,,\,\n_2]}
\label{curvM}
\ee 
is the curvature and obeys the algebraic identity
\be 
K_{[\m_1\ldots\m_{n+1}\,|\,\n_1]\n_2}=0\,.
\ee

\noindent The action (\ref{startingpoint}) and the curvature (\ref{curvM})
are invariant under the following gauge transformations
\begin{equation}
\d_{S,A}M_{\m_1\ldots\m_{n+1}}=
\partial_{[\m_1}S_{\m_2\ldots\m_n]\,|\,\m_{n+1}}
+\partial_{[\m_1}A_{\m_2\ldots\m_n]\m_{n+1}}
+\partial_{\m_{n+1}}A_{\m_2\ldots\m_n\m_1}
\label{invdejauge}
\end{equation}
where the gauge parameters $S_{\m_2\ldots\m_n\,|\,\m_{n+1}}$ and
$A_{\m_2\ldots\m_{n+1}}$ have the symmetries
\begin{center}
\begin{picture}(40,30)(0,-10)
\multiframe(0,10)(10.5,0){1}(10,10){\ft$2$}
\multiframe(10.5,10)(10.5,0){1}(17,10){\ft$n\!\!+\!\!1$}
\multiframe(0,-0.5)(10.5,0){1}(10,10){\ft$3$}
\put(4,-12){$\vdots$}
\multiframe(0,-13.5)(10.5,0){1}(10,12.5){$ $}
\multiframe(0,-24)(10.5,0){1}(10,10){\ft$n$}
\put(40,0){and}
\end{picture}
~~~~~~~~
\begin{picture}(50,30)(0,-10)
\multiframe(0,10)(10.5,0){1}(15,10){\ft$2$}
\multiframe(0,-0.5)(10.5,0){1}(15,10){\ft$3$}
\put(6,-12){$\vdots$}
\multiframe(0,-13.5)(10.5,0){1}(15,12.5){$ $}
\multiframe(0,-24)(10.5,0){1}(15,10){\tiny$n\!\!+\!\!1$}
\put(40,0){, respectively~.}
\end{picture}
\end{center}
\vspace*{10pt}
These gauge transformations are
accompanied by a chain of $n-1$ reducibilities on the gauge
parameters. 
These reducibilities read, with $1\leq i\leq n$ 
\ba
&&\stackrel{(i)}{S}_{\m_1\ldots\m_{n-i}\,|\,\m_{n-i+1}}=
\partial_{[\m_1}\stackrel{(i+1)}{S}_{\m_2\ldots\m_{n-i}]\,|\,\m_{n-i+1}}+
\nonumber \\
&&\frac{(n+1)}{(n-i+1)}\Big[\partial_{[\m_1}
\stackrel{(i+1)}{A}_{\m_2\ldots\m_{n-i}]\m_{n-i+1}}
+\partial_{\m_{n-i+1}}
\stackrel{(i+1)}{A}_{\m_2\ldots\m_{n-i}\m_1}\Big]\,,
\label{red1} \\
&&\stackrel{(i)}{A}_{\m_1\ldots\m_{n-i+1}}=\partial_{[\m_1}
\stackrel{(i+1)}{A}_{\m_2\ldots\m_{n-i+1}]}\,,
\label{red2}
\ea 
with the conventions that 
\ba
\stackrel{(1)}{S}_{\m_1\ldots\m_{n-1}\,|\,\m_n}&=&
S_{\m_1\ldots\m_{n-1}\,|\,\m_n}\,,\quad
\stackrel{(n)}{S}_\m\,=\, 0\,,\nn\\
\stackrel{(1)}{A}_{\m_1\ldots\m_n}&=&A_{\m_1\ldots\m_n}\,.\nn
\ea
The reducibility parameters at reducibility level $i$ have the
symmetry
\\
$\stackrel{(i+1)}{S}_{\m_1\ldots\m_{n-i-1}\,\mid\,\m_{n-i}}\simeq$
\footnotesize
\begin{picture}(40,30)(0,0)
\multiframe(0,10)(10.5,0){1}(20,10){$1$}
\multiframe(20.5,10)(10.5,0){1}(17,10){$n\!${\scriptsize{$-$}}$\!i$}
\multiframe(0,-0.5)(10.5,0){1}(20,10){$2$}
\put(7,-12){$\vdots$}
\multiframe(0,-24)(10.5,0){1}(20,10)
{\scriptsize{$n\!${\scriptsize{$-$}}$\!i${\scriptsize{$-$}}$\!\!$\scriptsize
{$1$}}}
\end{picture}\normalsize
and $\stackrel{(i+1)}{A}_{\m_1\ldots\m_{n-i}}\simeq$
\footnotesize
\begin{picture}(20,30)(0,0)
\multiframe(0,10)(10.5,0){1}(15,10){$1$}
\multiframe(0,-0.5)(10.5,0){1}(15,10){$2$}
\put(6.5,-12){$\vdots$}
\multiframe(0,-13.5)(10.5,0){1}(15,12.5){$ $}
\multiframe(0,-24)(10.5,0){1}(15,10){$n\!${\scriptsize{$-$}}$\!i$}
\end{picture}\normalsize
\vspace{1cm}.
\\
Note that $\stackrel{(n-1)}{S}_{\m\n}\,\simeq$ 
\footnotesize
\begin{picture}(20,20)(0,3)
\multiframe(0,0)(10.5,0){2}(10,10){$\m$}{$\n$}
\end{picture}
\normalsize.
These gauge transformations and reducibilities have already been
introduced and discussed in references
\cite{Curtright:1985,Aulakh:1986,Labastida:1986}. The problem of
investigating all the possible consistent couplings among the
fields $M_{\m_1\m_2\vert\m_3}$ will be treated in
\cite{Bekaert:2002uh}. Our theorem will provide a systematic tool for the
investigation of mixed symmetry type gauge field theories.

The number of physical degrees of freedom for the theory
(\ref{startingpoint}), (\ref{invdejauge}), is equal to \be
\frac{(D-2)!\,D\,(D-n-2)\,n}{(D-n-1)!\,(n+1)!}\,. \ee This number
is manifestly invariant under the exchange $n\leftrightarrow
D-n-2$ which corresponds to a Hodge duality transformation. This
confirms that the dimension for which the theory is dual to a
symmetric tensor is equal to $D=n+3$, which is also the critical
dimension for the theory to have local physical degrees of
freedom. The theory (\ref{startingpoint}), (\ref{invdejauge}) is
then dual to Pauli-Fierz's action (\ref{PF}) for $D=n+3$.
%
%
\section{$N$-complexes}\label{N-c}
%
%
The objective of the works presented in
\cite{Olver:1983,Dubois-Violette:1999,Dubois-Violette:2001} was to
construct complexes for irreducible tensor fields of mixed Young
symmetry type, thereby generalizing to some extent the calculus of
differential forms. This tool provides an elegant formulation of
symmetric tensor gauge fields and their Hodge duals (such as
differential form notation within electrodynamics).
%
\subsection{Young diagrams}\label{Young}
%
A {\bf Young diagram} $Y$ is a diagram which consists of a finite
number $S>0$ of columns of identical squares (referred to as the
{\bf cells}) of finite non-increasing lengths $l_1\geq l_2\geq
\ldots\geq l_S\geq 0$. For instance,
\ba
Y\equiv
\mbox{\footnotesize
\begin{picture}(38,15)(0,10)
\multiframe(10,14)(10.5,0){3}(10,10){}{}{}
\multiframe(10,3.5)(10.5,0){2}(10,10){}{}
\multiframe(10,-7)(10.5,0){2}(10,10){}{}
\multiframe(10,-17.5)(10.5,0){1}(10,10){}
\end{picture}\normalsize }
\nn\;\;\;.
\ea
\\
The total number of cells of the Young diagram $Y$ is
denoted by \be|Y|=\sum_{i=1}^S l_i\,.\ee
%
\subsubsection{Order relations}
%
For future reference, the subset ${\mathbb Y}^{(S)}$ of ${\mathbb
N}^S$ is defined by \be{\mathbb Y}^{(S)} \equiv
\{(n_1,\ldots,n_S)\in {\mathbb N}^S|\,n_1\geq n_2 \geq ...\geq
n_S\geq 0\}\,.\ee For two columns, the set ${\mathbb Y}^{(2)}$ can
be depicted as the following set of points in the plane ${\mathbb
R}^2$ : \be
\begin{array}{ccccc}
   &  &  &  & $\ldots$ \\
   &  &  & $$\stackrel{(3,3)}{\bullet}$$ & $\ldots$ \\
   &  & $$\stackrel{(2,2)}{\bullet}$$ & $$\stackrel{(3,2)}{\bullet}$$ &
$\ldots$ \\
   & $$\stackrel{(1,1)}{\bullet}$$ & $$\stackrel{(2,1)}{\bullet}$$ &
$$\stackrel{(3,1)}{\bullet}$$ & $\ldots$ \\
   $$\stackrel{(0,0)}{\bullet}$$ & $$\stackrel{(1,0)}{\bullet}$$ &
$$\stackrel{(2,0)}{\bullet}$$ & $$\stackrel{(3,0)}{\bullet}$$ & $\ldots$
\end{array}\ee

Let $Y$ be a diagram with $S$ columns of respective lengths $l_1$,
$l_2$, ..., $l_S$. If $Y_p$ is a well-defined Young diagram, then
$(l_1,l_2,\ldots,l_S)\in {\mathbb Y}^{(S)}$. Conversely, a Young
diagram $Y$ with $S$ columns is uniquely determined by the gift of
an element of ${\mathbb Y}^{(S)}$, and can therefore be labeled
unambiguously as $Y_{(l_1,l_2,\ldots,l_S)}^{(S)}$ ($l_S\neq 0$).
We denote\footnote{We will sometimes use the symbol ${\mathbb
Y}^{(S)}$ instead of $Y^{(S)}$.} by $Y^{(S)}$ the set of all Young
diagrams $Y_{(l_1,l_2,\ldots,l_S)}^{(S)}$ with at most $S$ columns
of respective length $0\leq l_S\leq l_{S-1}\leq\ldots\leq l_1\leq
D-1$. This identification between ${\mathbb Y}^{(S)}$ and
$Y^{(S)}$ suggests obvious definitions of sums and differences
of Young diagrams.

\noindent There is a natural definition of inclusion of Young
diagrams \be Y_{(m_1,\ldots,m_S)}^{(S)} \subset
Y_{(n_1,\ldots,n_S)}^{(S)}\,\Leftrightarrow\, m_1\leq
n_1\,,m_2\leq n_2\,,\ldots\,, m_S\leq n_S\,.\ee We can develop a
stronger definition of inclusion. Let $Y^{(S)}_{(m_1,\ldots,m_S)}$
and $Y^{(S)}_{(n_1,\ldots,n_S)}$ be two Young diagrams of
$Y^{(S)}$. We say that $Y^{(S)}_{(m_1,\ldots,m_S)}$ is {\bf
well-included} into $Y^{(S)}_{(n_1,\ldots,n_S)}$ if
$Y^{(S)}_{(m_1,\ldots,m_S)}\subset Y^{(S)}_{(n_1,\ldots,n_S)}$ and
$n_i-m_i\leq 1$ for all $i\in\{1,\ldots,S\}$. In other words, if
no column of the Young diagrams differs by more than a single box.
We denote this particular inclusion by $\Subset$, i.e. \be
Y_{(m_1,\ldots,m_S)}^{(S)} \Subset
Y_{(n_1,\ldots,n_S)}^{(S)}\,\Leftrightarrow\, m_i\leq n_i\leq
m_i+1\quad\forall i\in\{1,\ldots,S\}\,.\ee This new inclusion
suggests the following pictorial representation of ${\mathbb
Y}^{(S)}$ : \ba
\xymas{&&&&\ldots\\&&&\stackrel{(3,3)}{\bullet}\ar[r]\ar[ur]&\cdots
\\&&\stackrel{(2,2)}{\bullet} \ar[r]\ar[ur]&
\stackrel{(3,2)}{\bullet} \ar[u]\ar[r]\ar[ur]&\cdots \\
& \stackrel{(1,1)}{\bullet}\ar[r]\ar[ur]&
\stackrel{(2,1)}{\bullet} \ar[r]\ar[u]\ar[ur]&
\stackrel{(3,1)}{\bullet}\ar[u]\ar[r]\ar[ur]&\cdots \\
\stackrel{(0,0)}{\bullet} \ar[r]\ar[ur]&
\stackrel{(1,0)}{\bullet}\ar[u]\ar[r]\ar[ur]&
\stackrel{(2,0)}{\bullet}\ar[u]\ar[r]\ar[ur]
&\stackrel{(3,0)}{\bullet}\ar[u]\ar[r]\ar[ur] &\cdots
}\label{Ylatt}\ea where all the arrows represent maps $\Subset$.
This diagram is completely commutative.

The previous inclusions $\subset$ and $\Subset$ provide partial
order relations for ${\mathbb Y}^{(S)}$. The order is only partial
because all Young tableaux are not comparable.

\noindent
We now introduce a total order relation $\ll$
for ${\mathbb Y}^{(S)}$. \\ \noindent If $(m_1,\ldots,m_S)$ and
$(n_1,\ldots,n_S)$ belong to ${\mathbb Y}^{(S)}$, then \be
Y_{(m_1,\ldots,m_S)}^{(S)} \ll Y_{(n_1,\ldots,n_S)}^{(S)}
\,\,\Leftrightarrow \,\,\exists
K\in\{1,\ldots,S\} : \,\,\left\{\begin{array}{lll}m_i=n_i\,,
~~\forall
i\in\{1,\ldots, K\}\,, \\
m_{K+1}\leq n_{K+1}\,.\end{array}\right.  \label{Ygrading}\ee This
ordering simply provides the lexicographic ordering for ${\mathbb
Y}^{(S)}$.
%
\subsubsection{Maximal diagrams}
%
A sequence of ${\mathbb Y}^{(S)}$ which is of physical interest is
the \textbf{maximal sequence} denoted by $Y^S\equiv
(Y_p^S)_{p\in\mathbb N}$. This is defined as the naturally ordered
sequence of maximal diagrams\footnote{The subsequent notations for
maximal sequences are different from the ones of
\cite{Dubois-Violette:1999,Dubois-Violette:2001}. We have shifted
the upper index by one unit.} (the ordering is induced by the
inclusion of Young tableaux). \textbf{Maximal diagrams} are
diagrams with maximally filled rows, that is to say, Young
diagrams $Y^S_p$ with $p$ cells defined in the following manner:
we add cells to a row until it contains $S$ cells and then we
proceed in the same way with the row below, and continue until all
$p$ cells have been used. Consequently all rows except the last
one are of length $S$ and, if $r_p$ is the remainder of the
division of $p$ by $S$ ($r_p\equiv p\,\,\mbox{mod}\,S $) then the
last row of the Young diagram $Y^S_p$ will contain $r_p\leq S$
cells (if $r_p\not= 0$). For two columns ($S=2$) the maximal
sequence is represented as the following path in the plane
depicting ${\mathbb Y}^{(2)}$ :
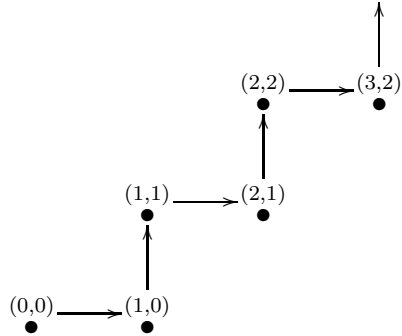
\ba\xymas{&&&\\&&\stackrel{(2,2)}{\bullet} \ar[r]&
\stackrel{(3,2)}{\bullet}\ar[u]\\
&\stackrel{(1,1)}{\bullet} \ar[r]&
\stackrel{(2,1)}{\bullet}\ar[u]\\
\stackrel{(0,0)}{\bullet} \ar[r]& \stackrel{(1,0)}{\bullet}\ar[u]
}\nn\ea Diagrams for which all rows have exactly $S$ cells are
called {\bf rectangular diagrams}. These are those represented by
the leftmost diagonal of the diagram ${\mathbb Y}^{(S)}$.
%
\subsubsection{Duality}
%
Let $Y_{(l_1,\ldots,l_S)}^{(S)}$ be a Young diagram in $Y^{(S)}$ and
 $I$  a non-empty subset of $\{1,\ldots,S\}$.

The diagram $D_{(\ell_1,\ldots,\ell_S)}^{(S)}$ with $S$ columns of
respective lengths  
\be
\ell_i\equiv\left\{\begin{array}{lll}l_i\,\quad&\mbox{if}\,\,i\not\in I\,,&
\\
D-l_i\,\quad&\mbox{if}\,\,i\in I\,,&\end{array}\right.
\label{ell}
\ee 
is, in general, \textit{not} a Young diagram. We
define the \textbf{dual Young diagram}
$\widetilde{Y}_{(\l_1,\ldots,\l_S)}^I\subset Y^{(S)}$ associated
to the set $I$ as the Young diagram obtained by reordering the
columns of $D_{(\ell_1,\ldots,\ell_S)}^{(S)}$.
In other words, its $i$-th column has length 
\be \l_i=\ell_{\pi(i)}\,,\quad 
\l_j\leq\l_i\,\,\mbox{for }i\leq j\,,\label{lambda}
\ee
where $\pi$ is a permutation of the elements of $\{1,\ldots,S\}$.
\subsubsection{Schur module}
%
Multilinear applications with a definite symmetry are associated
with a definite Young diagram \footnote{This set of definitions
essentially comes from \cite{Dubois-Violette:2001}.}. Let $V$ be a
finite-dimensional vector space of dimension $D$ and $V^\ast$ its
dual. The dual of the $n$-th tensor power $V^n$ of $V$ is
canonically identified with the space of multilinear forms:
$(V^n)^\ast\cong (V^\ast)^n$. Let $Y$ be a Young diagram and let
us consider that each of the $\vert Y\vert$ copies of $V^\ast$ in
the tensor product $(V^\ast)^{\vert Y\vert}$ is labeled by one
cell of $Y$. The {\bf Schur module}
 $V^Y$ is defined as the vector space of all multilinear forms
 $T$ in $(V^\ast)^{\vert Y\vert}$ such that :
\begin{quote}

$(i)$ $T$ is completely antisymmetric in the entries of each
column of $Y$,

$(ii)$ complete antisymmetrization of $T$ in the entries of a
column of $Y$ and another entry of $Y$ which is on the right-hand
side of the column vanishes.
\end{quote}
$V^Y$ is an irreducible subspace invariant for the action of
$GL(D,\mathbb R)$ on $V^{\vert Y\vert}$.

Let $Y$ be a Young diagram and $T$ an arbitrary multilinear
form in $(V^\ast)^{\vert Y\vert}$, 
one defines the multilinear form ${\cal Y}(T)\in (V^\ast)^{\vert Y\vert}$ 
by
\[
{\cal Y}(T)=T\circ{\cal A}\circ{\cal S}
\]
with
\[
{\cal A}=\sum_{c\in C}(-)^{\varepsilon(c)}c\,,\quad {\cal
S}=\sum_{r\in R} r
\]
where $C$ is the group of permutations which permute the entries
of each column, $\varepsilon(c)$ is the parity of the permutation
$c$, and $R$ is the group of permutations which permute the
entries of each row of $Y$. Any ${\cal Y} (T)\in V^Y$ and the
application ${\cal Y}$ of $V^{\vert Y\vert}$ satisfies the
condition ${\cal Y}^2=\lambda{\cal Y}$ for some number
$\lambda\not= 0$. Thus ${\bf Y} = \lambda^{-1}{\cal Y}$ is a
projection of $V^{\vert Y\vert}$ onto itself, i.e. ${\bf Y}^2={\bf
Y}$, with image Im$({\bf Y})=V^Y$. The projection ${\bf Y}$ will
be referred to as the {\bf Young symmetrizer} of the Young diagram
$Y$.
%
\subsection{Differential $N$-complex}
%
Let $(Y)=(Y_p)_{p\in \mathbb N}$ be a given sequence of Young
diagrams such that the number of cells of $Y_p$ is $p$, $\forall
p\in \mathbb N$. For each $p$, we assume that there is a single
shape $Y_p$ and $Y_p\subset Y_q$ for $p<q$. We define
$\Omega^p_{(Y)}({\cal M})$ as the vector space of smooth covariant
tensor fields of rank $p$ on the pseudo-Riemannian manifold ${\cal
M}$ which have the Young symmetry type $Y_p$ (i.e. their
components $T(x)$ belong to the Schur module $V^{Y_p}$ associated
to $Y_p$). More precisely they obey the identity ${\mathbf
Y}_p\,T(x)=T(x)$, $\forall x\in\cal M$, with ${\mathbf Y}_p$ the
Young symmetrizer on tensor of rank $p$ associated to the Young
symmetry $Y_p$. Let $\Omega_{(Y)}({\cal M})$ be the graded vector
space $\oplus_p\Omega^p_{(Y)}({\cal M})$ of irreducible tensor
fields on $\cal M$.

The exterior differential can then be generalized by setting
\cite{Dubois-Violette:1999,Dubois-Violette:2001}
\begin{equation}
d\equiv{\mathbf Y}_{p+1}\circ \partial : \Omega^p_{(Y)}({\cal
M})\rightarrow \Omega^{p+1}_{(Y)}({\cal M})\,,\label{defd}
\end{equation}
that is to say, first taking the partial derivative of the tensor
$T\in \Omega^p_{(Y)}({\cal M})$ and applying the Young symmetrizer
${\mathbf Y}_{p+1}$ to obtain a tensor in
$\Omega^{p+1}_{(Y)}({\cal M})$. Examples are provided in
subsection \ref{symt}.

Let us briefly mention that there are no $dx^\m$ in this
definition of the operator $d$. The operator $d$ is not nilpotent
in general, therefore $d$ does not always endow
$\Omega_{(Y)}({\cal M})$ with the structure of a standard
differential complex.

If we want to generalize the calculus of differential forms, we
have to use the extension of the structure of differential complex
with higher order of nilpotency. An {\bf $N$-complex} is defined
as a graded space $V=\oplus_i V_i$ equipped with an endomorphism
$d$ of degree $1$ that is nilpotent of order $N\in{\mathbb N}-
\{0,1\}$: $d^N=0$ \cite{Dubois-Violette:2000}. It is important to
stress that the operator $d$ is not necessarily a differential
because, in general, $d$ is neither nilpotent nor a derivative
(for instance, even if one defines a product in
$\Omega_{(Y)}({\cal M})$, the non-trivial projections affect the
Leibnitz rule).

A sufficient condition for $d$ to endow $\Omega_{(Y)}({\cal M})$
with the structure of an $N$-complex is that the number of columns
of any Young diagram be strictly smaller than $N$
\cite{Dubois-Violette:2001}:
\begin{lemma} Let $S$ be a non-vanishing integer and assume that the
sequence $(Y)$ is such that the number of columns of the Young
diagram $Y_p$ is strictly smaller than $S+1$ (i.e. $\leq S$) for
any $p\in \mathbb N$. Then the space $\Omega_{(Y)}({\cal M})$,
endowed with the operator $d$, is a $(S+1)$-complex.
\end{lemma}
\noindent Indeed, the condition $d^{S+1}\omega=0$ for all
$\omega\in\Omega_{(Y)}({\cal M})$ is fulfilled since the indices
in one column are antisymmetrized and $d^{S+1}\omega$ necessarily
involves at least two partial derivatives $\partial$ in one of the
columns (there are $S+1$ partial derivatives involved and a
maximum of $S$ columns).
\\
\\
\noindent\,\,{\bf Notation :} The space
$\Omega_{\left(Y^{(S)}\right)}({\cal M})$ is a $(S+1)$-complex
that we denote $\Omega_{(S)}({\cal M})$. The subcomplex
$\Omega_{Y_{(l_1,l_2,\ldots,l_S)}^{(S)}}({\cal M})$ is denoted by
$\Omega^{(l_1,l_2,\ldots,l_S)}_{(S)}({\cal M})$.

This complex $\Omega_{(S)}({\cal M})$ is the generalization of the
differential form complex $\Omega({\cal M})=\Omega_{(1)}({\cal
M})$ we are seeking for because each proper space is invariant
under the action of $GL(D,\mathbb R)$. For example, the previously
mentioned mixed symmetry type gauge field $M$ (\ref{YS}) belongs
to $\Omega^{(n,1)}_{(2)}({\cal M})$.
%
\subsection{Symmetric gauge tensors and maximal sequences}\label{symt}
%
A Young diagram sequence of interest in theories of spin $S\geq 1$
is the maximal sequence $Y^S=(Y^S_p)_{p\in\mathbb N}$
\cite{Dubois-Violette:1999,Dubois-Violette:2001}. This sequence is
defined as the sequence of diagrams with maximally filled rows
naturally ordered by the number $p$ of boxes.

\noindent\,\,{\bf Notation :} In order to simplify the notation, we
shall denote $\Omega^p_{(Y^S)}({\cal M})$ by $\Omega^p_S({\cal
M})$ and $\Omega_{(Y^S)}({\cal M})$ by $\Omega_S({\cal M})$.

\noindent If $D$ is the dimension of the manifold ${\cal M}$ then
the subcomplexes $\Omega^p_{S}({\cal M})$ with $p>SD$ are trivial
since, for these values of $p$, the Young diagrams $Y^S_p$ have at
least one column containing more than $D$ cells.
%
\subsubsection{Massless spin-one gauge field}
%
It is clear that $\Omega_1({\cal M})$ with the differential $d$ is
the usual complex $\Omega({\cal M})$ of differential forms on
${\cal M}$. The connection between the complex of differential
forms on ${\cal M}$ and the theory of classical $q$-form gauge
fields is well known. Indeed the subcomplex
\begin{equation}
\Omega^0({\cal M})\stackrel{d_0}{\rightarrow}\Omega^1({\cal
M})\stackrel{d_1}{\rightarrow}\ldots
\stackrel{d_{q-1}}{\rightarrow}\Omega^q({\cal
M})\stackrel{d_q}{\rightarrow}\Omega^{q+1}({\cal M})
\stackrel{d_{q+1}}{\rightarrow}\Omega^{q+2}({\cal
M})\label{complex}\end{equation} with $d_p\equiv d :
\Omega^p\rightarrow \Omega^{p+1}$, has the following
interpretation in terms of $q$-form gauge field $A_{[q]}$ theory.
The space $\Omega^{q+1}({\cal M})$ is the space which the field
strength $F_{[q+1]}$ lives in. The space $\Omega^{q+2}({\cal M})$
is the space of Hodge duals to magnetic sources $*J_m$ (at least
if we extend the space of ``smooth" $(q+2)$-forms to de Rham
currents) since $dF_{[q+1]}=(*J_m)_{[q+2]}$. If there is no
magnetic source, the field strength belongs to the kernel of
$d_{q+1}$. The Abelian gauge field $A_{[q]}$ belongs to
$\Omega^q({\cal M})$. The subspace Ker $d_q$ of $\Omega^q({\cal
M})$ is the space of pure gauge configurations (which are
physically irrelevant). The space $\Omega^{q-1}({\cal M})$ is the
space of infinitesimal gauge parameters $\Lambda_{[q-1]}$ and
$\Omega^{q-2}({\cal M})$ is the space of first reducibility
parameters $\Lambda_{[q-2]}$, etc. If the manifold ${\cal M}$ has
the topology of $\mathbb R^D$ then (\ref{complex}) is an exact
sequence.
%
\subsubsection{Massless spin-two gauge field}
%
As another example, we demonstrate the correspondence between some
Young diagrams in the maximal sequence with at most two columns
and their corresponding spaces in the differential $3$-complex
$\Omega_2({\cal M})$

\begin{table}[ht]
\begin{center}\begin{tabular}{|c|c|c|c|}
  \hline
  Young tableau & Vector space & Example & Components \\
  \hline
  \mbox{\footnotesize
\begin{picture}(38,15)(0,0)
\multiframe(10,0)(10.5,0){1}(10,10){}
\end{picture}\normalsize} & $\Omega^1_2({\cal M})$ & lin. diffeomorphism
parameter & $\xi_\m$ \\
&&&\\
  \mbox{\footnotesize
\begin{picture}(38,15)(0,0)
\multiframe(10,-1)(10.5,0){2}(10,10){}{}
\end{picture}\normalsize} & $\Omega^2_2({\cal M}) $ & graviton & $h_{\m\n}$ \\
&&&\\
  \mbox{\footnotesize
\begin{picture}(38,15)(0,0)
\multiframe(10,5)(10.5,0){2}(10,10){}{}
\multiframe(10,-5.5)(10.5,0){1}(10,10){}
\end{picture}\normalsize
}  & $\Omega^3_2({\cal M}) $ & mixed symmetry type field & $M_{\m\n\,|\,\r}$ \\
&&&\\
  \mbox{\footnotesize
\begin{picture}(38,15)(0,0)
\multiframe(10,3)(10.5,0){2}(10,10){}{}
\multiframe(10,-7.5)(10.5,0){2}(10,10){}{}
\end{picture}\normalsize} & $\Omega^4_2({\cal M}) $ & Riemann tensor &
$R_{\mu\nu\,\rho\sigma}$ \\
&&&\\
  \mbox{\footnotesize
\begin{picture}(38,15)(0,0)
\multiframe(10,9.5)(10.5,0){2}(10,10){}{}
\multiframe(10,-1)(10.5,0){2}(10,10){}{}
\multiframe(10,-11.5)(10.5,0){1}(10,10){}
\end{picture}\normalsize} & $\Omega^5_2({\cal M})$ & Bianchi identity &
 $\partial_{[\l}R_{\mu\nu]\,\rho\sigma}$ \\
&&&\\    \hline
\end{tabular}
\caption{Two-column maximal sequence and its physical relevance.}
\end{center}
\end{table}

The interest of $\Omega_2({\cal M})$ is its direct applicability
in free spin-two gauge theory. Indeed, in this case the analog of
the sequence (\ref{complex}) is
\begin{equation}
\Omega^1_2({\cal M})\stackrel{d}{\rightarrow} \Omega^2_2({\cal
M})\stackrel{d^2}{\rightarrow} \Omega^4_2({\cal
M})\stackrel{d}{\rightarrow}\Omega^5_2({\cal M})\label{complex2}
\end{equation} where $\Omega^1_2({\cal M})$ is
the space of covariant vector fields $\xi_\mu$ on ${\cal M}$,
$\Omega^2_2({\cal M})$ is the space of covariant rank $2$
symmetric tensor fields $h_{\mu\nu}$ on ${\cal M}$,
$\Omega^4_2({\cal M})$ the space of covariant tensor fields
$R_{\mu\nu\,\rho\sigma}$ of rank 4 having the symmetries of the
Riemann curvature tensor, and $\Omega^5_2({\cal M})$ is the space
of covariant tensor fields of degree 5 having the symmetries of
the left-hand side of the Bianchi II identity. The action of the
operator $d$, whose order of nilpotency is equal to 3, is written
explicitly in terms of components:
\begin{eqnarray}
(d\xi)_{\mu\nu}&=&\frac12(\partial_\mu \xi_\nu+\partial_\nu \xi_\mu) \\
(d^2h)_{\lambda\mu\rho\nu}&=&\frac14(\partial_\lambda\partial_\rho
h_{\mu\nu} +\partial_\mu\partial_\nu h_{\lambda\rho}-\partial_\mu
\partial_\rho h_{\lambda\nu} -\partial_\lambda\partial_\nu
h_{\mu\rho})\label{d2h}\\
(dR)_{\lambda\mu\nu\alpha\beta}&=&\frac13(\partial_\lambda
R_{\mu\nu\,\alpha\beta}+\partial_\mu
R_{\nu\lambda\,\alpha\beta}+\partial_\nu
R_{\lambda\mu\,\alpha\beta}).\label{dR}
\end{eqnarray}

The generalized $3$-complex $\Omega_2({\cal M})$ can be pictured
as the commutative diagram \ba\xymas{&&&&\cdots\\
&&&\Omega^6_2({\cal M})\ar[r]_{d}\ar[ur]^{d^2}&\cdots\\
&&\Omega^4_2({\cal M})\ar[r]_{d}\ar[ur]^{d^2}& \Omega^5_2({\cal M})\ar[u]_{d}\\
&\Omega^2_2({\cal M}) \ar[r]_{d}\ar[ur]^{d^2}& \Omega^3_2({\cal M})\ar[u]_{d}\\
\Omega^0_2({\cal M}) \ar[r]_{d}\ar[ur]^{d^2}& \Omega^1_2({\cal
M})\ar[u]_{d} }\label{maxseq}\ea In terms of this diagram, the
higher order nilpotency $d^3=0$ translates into the fact that (i)
if one takes a vertical arrow followed by a diagonal arrow, or
(ii) if a diagonal arrow is followed by a horizontal arrow, it
always maps to zero.
%
\subsection{Rectangular diagrams}\label{rectd}
%
The generalized cohomology \cite{Dubois-Violette:2000} of the
$N$-complex $\Omega_{N-1}({\cal M})$ is the family of graded
vector spaces $H_{(k)}(d)$ with $1\leq k \leq N-1$ defined by
$H_{(k)}(d)=\mbox{Ker}(d^k)/\mbox{Im}(d^{N-k})$. In general the
cohomology groups $H^p_{(k)}(d)$ are not empty, even when ${\cal
M}$ has a trivial topology. Nevertheless there exists a
generalization of the Poincar\'e lemma for $N$-complexes of
interest in gauge theories.

Let $Y^{S}$ be a maximal sequence of Young diagrams. The
(generalized) Poincar\'e lemma states that for ${\cal M}$ with the
topology of $\mathbb R^D$ the generalized
cohomology\footnote{Strictly speaking, the generalized Poincar\'e
lemma for rectangular diagrams was proved in
\cite{Dubois-Violette:1999,Dubois-Violette:2001} with an other
choice of convention where one first antisymmetrizes the columns.
This other choice is more convenient to prove
 the theorem in
\cite{Dubois-Violette:2001} but is inappropriate for considering
Hodge dualization properties. This explains our choice of
convention; still, as we will show later, the generalized
Poincar\'e lemma for rectangular diagrams remains true with the
definition (\ref{defd}).} of $d$ on tensors represented by
rectangular diagrams is empty in the space of maximal tensors
\cite{Olver:1983,Dubois-Violette:1999,Dubois-Violette:2001}.
\begin{proposition}(Generalized Poincar\'e lemma for rectangular
diagrams)\label{GenPoinrect}

\begin{itemize}
\item $H^0_{(k)}\left(\Omega_S(\mathbb R^D)\right)$ is the space of real
polynomial functions on $\mathbb R^D$ of degree strictly less than
$k$ ($1\leq k \leq N-1$) and
\item $H^{nS}_{(k)}\left(\Omega_S(\mathbb R^D)\right)=0$ $\forall n$ such that
$1\leq n\leq D-1$.
\end{itemize}
\end{proposition}

This is the first theorem of \cite{Dubois-Violette:2001}, the
proof of which is given therein. This theorem strengthens the
analogy between the two complexes (\ref{complex}) and
(\ref{complex2}) since it implies that (\ref{complex2}) is also an
exact sequence when ${\cal M}$ has a trivial topology.

Exactness at $\Omega^2_2({\cal M})$ means
$H^2_{(2)}\left(\Omega_2(\mathbb R^D)\right)=0$ and exactness at
$\Omega^4_2({\cal M})$ means $H^4_{(1)}\left(\Omega_2(\mathbb
R^D)\right)=0$. These properties have a physical interpretation in
terms of the linearized Bianchi identity II and gauge
transformations. Let $R_{\mu\nu\rho\sigma}$ be a tensor that is
antisymmetric in its two pairs of indices
$R_{\mu\nu\rho\sigma}=-R_{\nu\mu\rho\sigma}=-R_{\mu\nu\sigma\rho}$,
namely it has the symmetry of the Young diagram $~$ \footnotesize
\begin{picture}(60,15)(0,0)
\multiframe(0,3)(10.5,0){1}(10,10){}
\multiframe(0,-7.5)(10.5,0){1}(10,10){}\put(20,0){$\bigotimes$}
\multiframe(40,3)(10.5,0){1}(10,10){}
\multiframe(40,-7.5)(10.5,0){1}(10,10){}\end{picture}\normalsize .
This latter decomposes according to 
\ba
\begin{picture}(106,25)(0,0)
\multiframe(-80,10)(10.5,0){1}(10,10){}
\multiframe(-80,-0.5)(10.5,0){1}(10,10){}\put(-60,3){$\bigotimes$}
\multiframe(-38,10)(10.5,0){1}(10,10){}
\multiframe(-38,-0.5)(10.5,0){1}(10,10){} \put(-20,3){$=$}
\multiframe(0,10)(10.5,0){2}(10,10){}{}
\multiframe(0,-0.5)(10.5,0){2}(10,10){}{} \put(30,3){$\bigoplus$}
\multiframe(50,10)(10.5,0){2}(10,10){}{}
\multiframe(50,-0.5)(10.5,0){1}(10,10){}
\multiframe(50,-11)(10.5,0){1}(10,10){} \put(80,3){$\bigoplus$}
\multiframe(100,10)(10.5,0){1}(10,10){}
\multiframe(100,-0.5)(10.5,0){1}(10,10){}
\multiframe(100,-11)(10.5,0){1}(10,10){}
\multiframe(100,-21.5)(10.5,0){1}(10,10){}\put(120,3){.}
\end{picture}
\label{Yougdec}\\\nonumber
\ea 
If we impose the condition that $R$
obeys the first Bianchi identity (\ref{BianchiI}), we eliminate
the last two terms in its decomposition (\ref{Yougdec}) hence the
tensor $R$ has the symmetries of the Riemann tensor and belongs to
$\Omega^4_2({\cal M})$. Furthermore, from (\ref{dR}) it is
apparent that the linearized second Bianchi identity
(\ref{BianchiII}) for $R$ reads $dR=0$. As the Riemann tensor has
the symmetries of a rectangular diagram, we obtain $R=d^2h$ with
$h\in\Omega^2_2({\cal M})$  from the exactness of the sequence
(\ref{complex2}). This means that $R$ is effectively the
linearized Riemann tensor associated to the spin-two field $h$, as
can be directly seen from (\ref{d2h}). However, the definition of
the metric fluctuation $h$ is not unique : the gauge field
$h+\delta h$ is physically equivalent to $h$ if it does not affect
the physical linearized Riemann tensor, i.e. $d^2(\delta h)=0$.
Since the sequence (\ref{complex}) is exact we find : $\delta
h=d\xi$ with $\xi\in\Omega^1_2({\cal M})$. As a result we recover
the standard gauge transformations (\ref{2gauge}).
%
\subsection{Multiforms, Hodge duality and trace operators}\label{Hodged}
%
A good mathematical understanding of the gauge structure of free
symmetric tensor gauge field theories is provided by the maximal
sequence and the vanishing of the rectangular diagrams cohomology.
However, several new mathematical ingredients are needed as well
as an extension of Proposition \ref{GenPoinrect} to capture their
dynamics. A useful new ingredient is the obvious generalization of
Hodge's duality for $\Omega_S(\mathbb R^D)$, which is obtained by
contracting the columns with the epsilon tensor
$\varepsilon^{\mu_1\dots\mu_D}$ of ${\cal M}$ and lowering the
indices with the Minkowskian metric. For rank $S$ symmetric tensor
gauge theories there are $S$ different Hodge operations since the
corresponding diagrams may contain up to $S$ columns. A simple but
important point to note is the following: generically the Hodge
duality is not an internal operation in the space $\Omega_S({\cal
M})$. For this reason, we define a new space of tensors in the
next subsection.
%
\subsubsection{Multiforms}
%
A key ingredient is the graded tensor product of $C^\infty({\cal
M})$ with $S$ copies of the exterior algebra $\Lambda{\mathbb
R}^{D*}$ where ${\mathbb R}^{D*}$ is the dual space of basis $d_i
x^\mu$ ($1\leq i\leq S$, thus there are $S$ times $D$ of them).
Elements of this space will be referred to as {\bf multiforms}
\cite{Dubois-Violette:2001}. They are sums of products of the
generators $d_ix^\mu$ with smooth functions on $\cal M$. The
components of a multiform define a tensor with the symmetry
properties of the product of $S$ columns
\vspace{0.3cm}\begin{center}\footnotesize
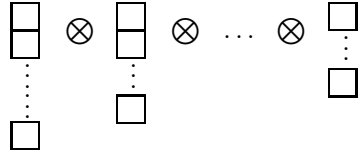
\begin{picture}(0,37)(0,-24)
\multiframe(-40,10)(10.5,0){1}(10,10){}
\multiframe(-40,-0.5)(10.5,0){1}(10,10){}
\put(-35.5,-12){$\vdots$}\put(-35.5,-23.5){$\vdots$}
\multiframe(-40,-35)(10.5,0){1}(10,10){} \put(-20,7){$\bigotimes$}
\multiframe(0,10)(10.5,0){1}(10,10){}
\multiframe(0,-0.5)(10.5,0){1}(10,10){} \put(4.5,-12){$\vdots$}
\multiframe(0,-25)(10.5,0){1}(10,10){} \put(20,7){$\bigotimes$}
\put(40,7){$\ldots$} \put(60,7){$\bigotimes$}
\multiframe(80,10)(10.5,0){1}(10,10){}\put(85,-1.5){$\vdots$}
\multiframe(80,-15)(10.5,0){1}(10,10){}
\end{picture}\normalsize\end{center}\vspace{0.3cm}

\noindent\,\,{\bf Notation :} We shall denote this multigraded
space $\big(\otimes^{S}\Lambda({\mathbb R}^{D*}\big)\otimes
C^\infty({\cal M})$ by $\Omega_{[S]}({\cal M})$. The subspace
$\Omega^{l_1,l_2,\ldots,l_S}_{[S]}({\cal M})$ is defined as the
space of multiforms whose components have the symmetry properties
of the diagram $D_{l_1,l_2,\ldots,l_S}:=\bigotimes\limits_{i=1}^S
Y^{(1)}_{(l_i)}$ which represents the above product of $S$ columns
with respective lengths $l_1$, $l_2$, ..., $l_S$.

The tensor field
$\a_{[\mu^1_1\ldots\mu^1_{l_1}]\ldots[\mu^S_1\ldots\mu^S_{l_S}]}(x)$
defines a multiform $\a\in\Omega^{l_1,\ldots,l_S}_{[S]}({\cal M})$
which explicitly reads \be
\a=\a_{[\mu^1_1\ldots\mu^1_{l_1}]\ldots[\mu^S_1\ldots\mu^S_{l_S}]}(x)\,\,
d_1x^{\mu^1_1}\wedge\ldots\wedge
d_1x^{\mu^1_{l_1}}\,\ldots\,d_Sx^{\mu^S_1}\wedge\ldots\wedge d_S
x^{\mu^S_{l_S}} \,.\label{multif}\ee In the sequel, when we refer
to the multiform $\a$ we will speak either of (\ref{multif}) or of
its components. More accurately, we will identify
$\Omega_{[S]}({\cal M})$ with the space of the smooth tensor field
components.

We endow $\Omega_{[S]}({\cal M})$ with the structure of a
(multi)complex by defining $S$ anticommuting differentials \be
d_i : \Omega^{l_1,\ldots,l_i,\ldots,l_S}_{\,\,[S]}({\cal
M})\rightarrow\Omega^{l_1,\ldots,l_i+1,\ldots,l_S}_{\,\,[S]}({\cal
M})\,,\quad 1\leq i\leq S \,,\ee defined by adding a box
containing the partial derivative in the $i$-th column. For
instance, $d_2$ acting on the previous diagrammatic example is
\vspace{0.3cm}
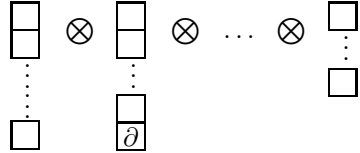
\begin{center}\footnotesize
\begin{picture}(0,37)(0,-24)
\multiframe(-40,10)(10.5,0){1}(10,10){}
\multiframe(-40,-0.5)(10.5,0){1}(10,10){}
\put(-35.5,-12){$\vdots$}\put(-35.5,-23.5){$\vdots$}
\multiframe(-40,-35)(10.5,0){1}(10,10){} \put(-20,7){$\bigotimes$}
\multiframe(0,10)(10.5,0){1}(10,10){}
\multiframe(0,-0.5)(10.5,0){1}(10,10){} \put(4.5,-12){$\vdots$}
\multiframe(0,-25)(10.5,0){1}(10,10){}
\multiframe(0,-35.5)(10.5,0){1}(10,10){$\partial$}
\put(20,7){$\bigotimes$} \put(40,7){$\ldots$}
\put(60,7){$\bigotimes$}
\multiframe(80,10)(10.5,0){1}(10,10){}\put(85,-1.5){$\vdots$}
\multiframe(80,-15)(10.5,0){1}(10,10){}
\end{picture}\normalsize\end{center}
\vspace{0.3cm}

\noindent\,\,{\bf Summary of notations :} The multicomplex
$\Omega_{[S]}({\cal M})$ is the subspace of $S$-uple multiforms.
It is the direct sum of subcomplexes
$\Omega^{l_1,l_2,\ldots,l_S}_{[S]}({\cal M})$. The space
$\Omega_{(S)}({\cal M})$ is the $(S+1)$-complex of tensors
represented by Young tableaux with at most $S$ columns. It is the
direct sum of subcomplexes
$\Omega^{(l_1,l_2,\ldots,l_S)}_{(S)}({\cal M})$. The space
$\Omega_S({\cal M})=\oplus_p\Omega^p_S({\cal M})$ is the space of
maximal tensors. Thus we have the chain of inclusions
$\Omega_{S}({\cal M})\subset \Omega_{(S)}({\cal M})
\subset\Omega_{[S]}({\cal M})$.
%
\subsubsection{Hodge and trace operators}
%
We introduce the following notation for the $S$ possible Hodge
dual definitions \be
*_i:\Omega^{l_1,\ldots,l_i,\ldots,l_S}_{\,\,[S]}({\cal
M})\rightarrow\Omega^{l_1,\ldots,D-l_i,\ldots,l_S}_{\,\,[S]}({\cal
M})\,,\quad 1\leq i\leq S \,.\ee The operator $*_i$ is defined as
the action of the Hodge operator on the indices of the $i$-th
column. To remain in the space of covariant tensors requires the
use of the flat metric to lower down indices.

Using the metric, another simple operation that can be defined is
the trace. The convention is that we always take the trace over
indices in two different columns, say the $i$-th and $j$-th. We
denote this operation by\be
\mbox{Tr}_{ij}:\Omega^{l_1,\ldots,l_i,\ldots,l_j,\ldots,l_S}_{\,\,[S]}({\cal
M})\rightarrow\Omega^{l_1,\ldots,l_i-1,\ldots,l_j-1,\ldots,l_S}_{\,\,[S]}({\cal
M})\,,\quad i\neq j \,.\ee The Schur module definition (see
subsection \ref{Young}) gives the necessary and sufficient set of
conditions for a (covariant) tensor $T_{\m_1\m_2\ldots\m_p}(x)$ of
rank $p$ to be in the irreducible representation of $GL(D,{\mathbb
R})$ associated with the Young diagram $Y$ (with $|Y|=p$). Each
index of $T_{\m_1\m_2\ldots\m_p}(x)$ is placed in one box of $Y$.
The set of conditions is the following :
\begin{quote}

$(i)$ $T_{\m_1\m_2\ldots\m_p}(x)$ is completely antisymmetric in
the entries of each column of $Y$,

$(ii)$ complete antisymmetrization of $T_{\m_1\m_2\ldots\m_p}(x)$
in the entries of a column of $Y$ and another entry of $Y$ which
is on the right of the column, vanishes.

\end{quote}
Using the previous definitions of multiforms, Hodge dual and trace
operators, this set of conditions gives
\begin{proposition}(Schur module)

Let $\a$ be a multiform in $\Omega^{l_1,\ldots,l_S}_{[S]}({\cal
M})$. If \ba l_j\leq l_i<D\,,\quad\forall\,
i,j\in\{1,\ldots,S\}:\,\,i\leq j\,,\nn\ea then one obtains the
equivalence\ba \mbox{Tr}_{ij}\,\{\,*_i\,\a\,\}\,=\,0\quad\forall\,
i,j:\,\, 1\leq i<j\leq S \quad\Longleftrightarrow\quad \a\in
\Omega^{(l_1,\ldots,l_S)}_{(S)}({\cal M})\,.\nn\ea\label{Schurm}
\end{proposition}
Indeed, condition (i) is satisfied since $\a$ is a multiform.
Condition (ii) is simply rewritten in terms of tracelessness
conditions.

Another useful property, which generalizes the derivation followed
in the chain of equations (\ref{*R})-(\ref{p*R}), is for any
$i,j\in\{1,\ldots,S\}$ \be
\bullet\quad\left\{\begin{array}{lll}\mbox{Tr}_{ij}\,\a\,=\,0&& \\
d_i \a\,=\,0&&\end{array}\right.\quad\Longrightarrow\quad
d_j\,(*_j\,\a)\,=\,0\,.\label{makeuse}\ee

The following property on powers of the trace operator will also
be useful later on. We state it as
\begin{proposition}\label{tracepower}
Let $\a\in\Omega^{l_1,\ldots,l_S}_{[S]}({\cal M})$ be a
multiform. For any $m\in\mathbb N$ such that $0\leq m\leq min(D-l_i,D-l_j$), 
one has the equivalence \ba
\big(\mbox{Tr}_{ij}\big)^m\{*_i*_j\a\}=0\quad\quad\Longleftrightarrow\quad
\big(\mbox{Tr}_{ij}\big)^{m+l_i+l_j-D}\,\{\,\a\,\}=0\,.\nn\ea
\end{proposition}
\proof{The proof of the proposition is inductive, the induction parameter 
being the number of traces, and is mainly based on the rule
for contractions of epsilon tensors.

\begin{description}

\item[$\underline{\,\bf\Rightarrow:}$] We start the proof of the
necessity by a \underline{preliminary lemma: }

For any given integer
$p\in{\mathbb N}$, \be\left.\begin{array}{lll}
\big(\mbox{Tr}_{ij}\big)^{D-l_i-p}\{*_i*_j\a\}=0&& \\
\big(\mbox{Tr}_{ij}\big)^{l_j-n+1}\{\a\}=0\,,\,\,\,\forall n\geq p
&&\end{array}\right\}\quad\Rightarrow\quad
\big(\mbox{Tr}_{ij}\big)^{l_j-p}\{\a\}=0\,.\label{bogol}\ee This
is true because it can be checked explicitly that
$\big(\mbox{Tr}_{ij}\big)^{D-l_i-p}\{*_i*_j\a\}$ is equal to a sum
of terms proportional to $\big(\mbox{Tr}_{ij}\big)^{l_j-k}\{\a\}$
for all $k\geq p$. The second hypothesis says that these last
terms vanish for $k\geq p+1$. As a result,
$\big(\mbox{Tr}_{ij}\big)^{D-l_i-p}\{*_i*_j\a\}$
$\propto\big(\mbox{Tr}_{ij}\big)^{l_j-p}\{\a\}$.
Therefore the vanishing of
$\big(\mbox{Tr}_{ij}\big)^{D-l_i-p}\{*_i*_j\a\}$ implies the
vanishing of $\big(\mbox{Tr}_{ij}\big)^{l_j-p}\{\a\}$.

Now that this preliminary lemma is given, we can turn back to our 
inductive proof.\\ \noindent
The induction hypothesis $I_m$ is the following : 
\ba
\big(\mbox{Tr}_{ij}\big)^m\{*_i*_j\a\}=0\quad
{\Rightarrow}\quad
\big(\mbox{Tr}_{ij}\big)^{m+l_i+l_j-D}\,\{\,\a\,\}=0\,.
\label{lftrt}
\ea
The starting point of the induction is $I_{D-l_i+1}$ [considering without loss 
of generality that $D-l_i=min(D-l_i,D-l_j)$] which is obviously true 
since in this case where $m=D-l_i+1$, both traces in (\ref{lftrt}) vanish.
What we have to show now is that, if $I_n$ is true $\forall ~n\geq m$,
then $I_{m-1}$ is also true. 

It is obvious that
$\big(\mbox{Tr}_{ij}\big)^{m-1}\{*_i*_j\a\}=0$ implies that
$\big(\mbox{Tr}_{ij}\big)^n\{*_i*_j\a\}=0$ for all $n\geq m-1$.
The induction hypothesis thus implies that
$\big(\mbox{Tr}_{ij}\big)^{n+l_i+l_j-D}\,\{\,\a\,\}=0$ for all
$n\geq m$. Together with
$\big(\mbox{Tr}_{ij}\big)^{m-1}\{*_i*_j\a\}=0$ and the help of the
lemma (\ref{bogol}), we eventually obtain \\ \noindent
$\big(\mbox{Tr}_{ij}\big)^{m-1+l_i+l_j-D}\,\{\,\a\,\}=0$, which
ends the proof of the induction hypothesis.

\item[$\underline{\,\bf\Leftarrow:}$] In this case, the
sufficiency is a consequence of the necessity. In other words,
since we proved that the implication $I_m$ is valid from the left
to the right in (\ref{lftrt}) we will show that then, it is also
valid from the right to the left.

Indeed, the relation $*_i*_j(*_i*_j\a)=\pm\a$ allows to write \be
\big(\mbox{Tr}_{ij}\big)^{m+l_i+l_j-D}\,\{\,\a\,\}=
\pm\big(\mbox{Tr}_{ij}\big)^{m+(l_i-D)+(l_j-D)+D}\,\{\,*_i*_j(*_i*_j\a)\,\}\,.\label{whoknows}\ee
The (proven) implication $I_m$ of Proposition \ref{tracepower}
applied to the multiform
$*_i*_j\a\in\Omega^{l_1,\ldots,D-l_i,\ldots,D-l_j,\ldots,l_S}_{(S)}({\cal
M})$ is
\be
\big(\mbox{Tr}_{ij}\big)^{m+(l_i-D)+(l_j-D)+D}\,\{\,*_i*_j(*_i*_j\a)\,\}=0
\;\Rightarrow\;
\big(\mbox{Tr}_{ij}\big)^{m}\,\{\,\a\,\}=0\,.\ee Combined with the
relation (\ref{whoknows}), the previous implication is precisely
the (reversed) implication in Proposition \ref{tracepower}.

\end{description}
}
%
\subsection{Generalized nilpotency}
%
Let $Y_p$ be well-included\footnote{See subsection \ref{Young}}
into $Y_{p+q}$, that is $Y_p\Subset Y_{p+q}$. Let $I$ be the
subset of $\{1,2, \dots, S\}$ containing the $q$ elements ($\# I =
q$) corresponding to the difference between $Y_{p+q}$ and $Y_p$.
We ``generalize" the definition (\ref{defd}) by introducing the
differential operators $d^I$ as follows (see also
\cite{Olver:1983,Olver:1987})
\begin{equation}  d^I\equiv c^p_I\,{\mathbf Y}_{p+q}\circ \big(\prod_{i\in I}
\partial_i\big) : \Omega^p_{(Y)}({\cal M})\rightarrow
\Omega^{p+q}_{(Y)}({\cal M})
\end{equation}
where $\partial_i$ indicates that the index corresponding to this
partial derivative is placed at the bottom of the $i$-th column
and $c^p_I$ are normalization factors so that we have strict
equalities in the next Proposition \ref{commutprop} \footnote{The
precise expression for the constants $c^p_I$ was obtained in
\cite{Olver:1983}.}. When $I$ contains only one element ($q = 1$)
we recover the definition (\ref{defd}) of $d$. The tensor $d^IY_p$
will be represented by the Young diagram $Y_{p+q}$ where we place
a partial derivative symbol $\partial$ in the $q$ boxes which do
not belong to the subdiagram $Y_p\Subset Y_{p+q}$.

The product of operators $d^I$ is commutative : $d^I\circ
d^J=d^J\circ d^I$ for all $I,J\subset\{1,\ldots,S\}$ ($\#I=q$,
$\#J=r$) such that the product maps to a well-defined Young
diagram $Y_{p+q+r}$. The following proposition gathers all these
properties
\begin{proposition}\label{commutprop}
Let $I$ and $J$ be two subsets of $\{1,\ldots,S\}$. Let
$\a$ be an irreducible tensor belonging to $\Omega_{(2)}({\cal M})$. The
following properties are satisfied\footnote{According to the
terminology of \cite{Olver:1983}, these properties mean that the
set $Y^{(S)}$ is endowed with the structure of
hypercomplex by means of the maps $d^I$.}
\begin{itemize}
  \item If $I\cap J=\phi$, then $(d^I\circ d^J)\a\,=\,d^{\,I\cup
  J}\a$. Therefore, $d^I\a\,=\,d^{\#I}\a$
  \item If $I\cap J\neq\phi$, then $(d^I\circ d^J)\a=0$.
\end{itemize}
\end{proposition}
Proposition \ref{commutprop} is proved in \cite{Olver:1983}. The
last property states that the product of $d^I$ and $d^J$
identically vanishes if it is represented by a diagram $Y_{p+q+r}$
with at least one column containing two partial derivatives.
Proposition \ref{commutprop} proves that the operator $d^I$
provides the most general non-trivial way of applying partial
derivatives in $\Omega_{(S)}({\cal M})$.

Proposition \ref{commutprop} is also helpful because it makes
contact with the definition (\ref{defd}) in that the operator
$d^I$ can be identified, up to a constant factor, with the
(non-trivial) $\# I$-th power of the operator $d$. Despite this
identification, we frequently use the notation $d^I$ because it
contains more information than the notation $d^{\# I}$.

The space $\Omega_{(2)}({\cal M})$ can be pictured analogously to
the representation (\ref{Ylatt}) of the set of Young diagrams
$Y^{(2)}$
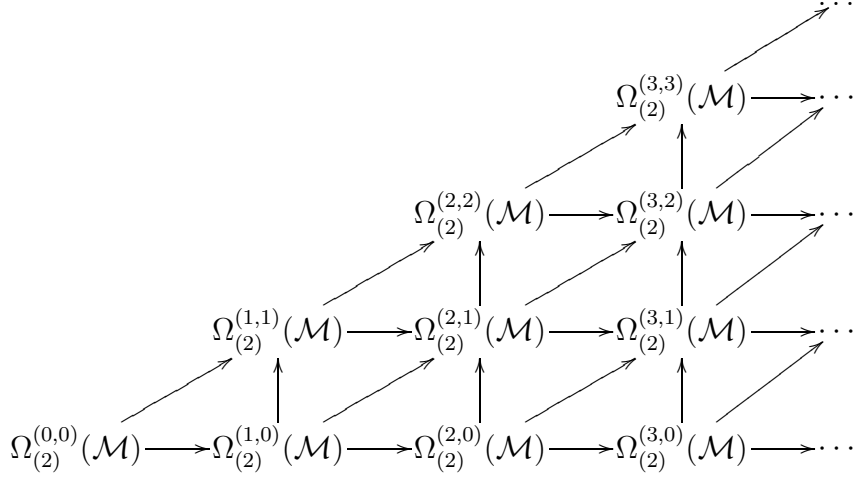
\ba\xymas{&&&&\ldots\\&&&\Omega^{(3,3)}_{(2)}({\cal
M})\ar[r]\ar[ur]&\cdots
\\&&\Omega^{(2,2)}_{(2)}({\cal M}) \ar[r]\ar[ur]&
\Omega^{(3,2)}_{(2)}({\cal M}) \ar[u]\ar[r]\ar[ur]&\cdots \\
& \Omega^{(1,1)}_{(2)}({\cal M}) \ar[r]\ar[ur]&
\Omega^{(2,1)}_{(2)}({\cal M}) \ar[r]\ar[u]\ar[ur]&
\Omega^{(3,1)}_{(2)}({\cal M})\ar[u]\ar[r]\ar[ur]&\cdots \\
\Omega^{(0,0)}_{(2)}({\cal M}) \ar[r]\ar[ur]&
\Omega^{(1,0)}_{(2)}({\cal M})\ar[u]\ar[r]\ar[ur]&
\Omega^{(2,0)}_{(2)}({\cal M})\ar[u]\ar[r]\ar[ur]
&\Omega^{(3,0)}_{(2)}({\cal M})\ar[u]\ar[r]\ar[ur] &\cdots
}\nn\ea\vspace{.5cm}

\noindent From the previous discussions, the definitions of the
arrows should be clear:
\begin{itemize}
  \item[$\rightarrow$ :] Horizontal arrows are maps $d= d^{\{1\}}$.
  \item[$\uparrow$ :] Vertical arrows are maps $d= d^{\{2\}}$.
  \item[$\nearrow$ :] Diagonal arrows are maps $d^2= d^{\{1,2\}}$.
\end{itemize}
Proposition \ref{commutprop} translates in terms of this diagram
into the fact that
\begin{itemize}
  \item this diagram is completely commutative, and
  \item the composition of any two arrows with at least one common
  direction maps to zero identically.
\end{itemize}
Of course, these diagrammatic properties hold for arbitrary $S$
(the corresponding picture would be simply a higher-dimensional
generalization since ${\mathbb Y}^{(S)}\subset {\mathbb R}^S$).
%
\section{Linearized gravity field equations}\label{fieldequs}
%
From now on, we will restrict ourselves to the case of linearized
gravity, i.e. rank-$2$ symmetric gauge fields. There are two
possible Hodge operations, denoted by $*$, acting on the first
column if it is written on the left, and on the second column if
it is written on the right. Since we are no longer restricted to
maximal Young diagrams the notation $d$ is ambiguous (we do not
know a priori on which Young symmetry type we should project in
the definition (\ref{defd})). Instead we use the above mentioned
differentials $d_i$ of multiform theory. There are only two of
these in the case of linearized gravity: $d_1$ called the (left)
differential, denoted by $d_L$, and $d_2$, the (right)
differential, denoted by $d_R$. With these differentials it is
possible to rewrite (\ref{d*R'}) in the compact form
$d_L*R=0=d_RR*$. The second Bianchi identity reads $d_L R=0=d_RR$.

The convention that we use is to take the trace over indices in
the first row, using the flat background metric $\eta_{\mu\nu}$.
We denote this operation by Tr (which is $\mbox{Tr}_{12}$
according to the definition given in the previous section). In
this notation the Einstein equation (\ref{Einstein}) takes the
form Tr$R=0$, while the first Bianchi identity (\ref{BianchiI})
reads Tr$*R=0$. From Proposition \ref{Schurm}, it is clear that
the following property holds: let $B$ be a ``biform" in
$\Omega^{p,q}_{[2]}({\cal M})$ which means $B$ is a tensor with
symmetry 
\be 
\begin{picture}(20,37)(0,-24)
\multiframe(0,10)(10.5,0){1}(10,10){$1$}
\multiframe(0,-0.5)(10.5,0){1}(10,10){$2$}
\put(4.5,-12){$\vdots$}\put(4.5,-23.5){$\vdots$}
\multiframe(0,-35)(10.5,0){1}(10,10){$p$} \put(20,7){$\bigotimes$}
\multiframe(40,10)(10.5,0){1}(10,10){$1$} \put(45,-1.5){$\vdots$}
\multiframe(40,-15)(10.5,0){1}(10,10){$q$} \put(60,-15){,}
\end{picture}
\nonumber
\ee
\\
then, $B$ obeys the (first) ``Bianchi identity"
\be
\mbox{Tr}(*B)=0
\ee 
if and only if $B\in
\Omega^{(p,q)}_{(2)}({\cal M})$. 
This is pictorially described by
the diagram 
\be
\vspace{.2cm}
\begin{picture}(20,37)(0,-24)
\multiframe(0,10)(10.5,0){1}(10,10){}
\multiframe(0,-0.5)(10.5,0){1}(10,10){}
\put(4.5,-12){$\vdots$}\put(4.5,-23.5){$\vdots$}
\multiframe(0,-35)(10.5,0){1}(10,10){}
\multiframe(10.5,10)(10.5,0){1}(10,10){} \put(15.5,-1.5){$\vdots$}
\multiframe(10.5,-16)(10.5,0){1}(10,10){}
\end{picture}
\nonumber
\ee 
that is, the two columns of the product are attached together.

With all the new artillery introduced in the previous section, it
becomes easier to extend the concept of electric-magnetic duality
for linearized gravity. First of all we emphasize the analogy
between the Bianchi identities and the field equations by
rewriting them respectively as
\be
\left\{\begin{array}{lll}\mbox{Tr}*R=0&& \\
d_LR=0=d_RR&&\end{array}\right. ,\label{BianchiIII}
\ee 
and
\be
\left\{\begin{array}{lll}\mbox{Tr}R=0&& \\
d_L(*R)=0=d_R(R*)&&\end{array}\right. ,\label{fieldequ}
\ee 
where
$R_{\mu\nu\,\rho\sigma}\equiv\mbox{ \footnotesize
\begin{picture}(50,15)(0,0)
\multiframe(-3,6.5)(10.5,0){1}(10,10){$\mu$}
\multiframe(-3,-4)(10.5,0){1}(10,10){$\nu$}\put(15,1){$\bigotimes$}
\multiframe(31,6.5)(10.5,0){1}(10,10){$\rho$}
\multiframe(31,-4)(10.5,0){1}(10,10){$\sigma$}
\end{picture}}$. We recall that $d_L(*R)=0=d_R(R*)$ was obtained in section
\ref{Lingravi} by
using the second Bianchi identity.

As discussed in subsection \ref{Hodged}, the first Bianchi
identity implies that $R$ effectively has the symmetry properties
of the Riemann tensor, i.e. $R_{\mu\nu\,\rho\sigma}\equiv\mbox{
\footnotesize
\begin{picture}(25,15)(0,0)
\multiframe(-3,6.5)(10.5,0){1}(10,10){$\mu$}
\multiframe(-3,-4)(10.5,0){1}(10,10){$\nu$}
\multiframe(7.5,6.5)(10.5,0){1}(10,10){$\rho$}
\multiframe(7.5,-4)(10.5,0){1}(10,10){$\sigma$}
\end{picture}}$. Using this symmetry
property the two equations $d_LR=0=d_RR$ can now be rewritten as
the single equation $dR=0$. Therefore, if the manifold ${\cal M}$
is of trivial topology then, for a given multiform $R\in
\Omega^{2,2}_{[2]}({\cal M})$, one obtains the equivalence
\be
\left\{\begin{array}{lll}\mbox{Tr}*R=0&& \\
d_LR=0=d_RR&&\end{array}\right.\Leftrightarrow\quad
\left\{\begin{array}{lll}R=d^2h &&\\
h\in\Omega^2_2({\cal M})&&\end{array}\right. \,,
\label{Bibi}
\ee
due to Proposition \ref{GenPoinrect} and Proposition \ref{Schurm}.
%
\subsection{Dual linearized Riemann tensor}
%
By Proposition \ref{Schurm}, the (vacuum) Einstein equation
Tr$R=0$ can then be translated into the assertion that the dual of
the Riemann tensor has (on-shell) the symmetries of a diagram
$(D-2,2)$, in other words $*R\in\Omega^{(D-2,2)}_{(2)}({\cal M})$.
As explained in subsection \ref{duallingravi}, the second Bianchi
identity $d_R R=0$, together with the linearized Einstein
equations, implies the equation $d_L*R=0$. Furthermore the second
Bianchi identity $d_R R=0$ is equivalent to $d_R*R=0$, therefore
we have the equivalence
\be
\left\{\begin{array}{lll}\mbox{Tr}R=0&& \\
d_RR=0&&\end{array}\right.\Leftrightarrow\quad
\left\{\begin{array}{lll}*R\in\Omega^{(D-2,2)}_{(2)}({\cal M}) &&\\
d_L*R=0=d_R*R &&\end{array}\right. \,.
\ee 
In addition
$d_L*R=0=d_R*R$ now implies $*R=d^2\tilde{h}$ (where we denote the
non-ambiguous product $d_L\,d_R$ by $d^2$). The tensor field
$\tilde{h}\in \Omega^{(D-3,1)}_{(2)}({\cal M})$ is the dual gauge
field of $h$ obtained in (\ref{dualcurva}). This property
(\ref{dualcurva}), which holds for manifolds $\cal M$ with the
topology of ${\mathbb R}^D$, is a direct application of Corollary
\ref{coro} of the generalized Poincar\'e lemma given in the
following section; we anticipate this result here in order to
motivate the theorem by using a specific example. We have an
equivalence analogous to (\ref{Bibi}),
\be
\left\{\begin{array}{lll}\mbox{Tr}R=0&& \\
d_L*R=0=d_R*R&&\end{array}\right.\Leftrightarrow\quad
\left\{\begin{array}{lll}*R=d^2\tilde{h} &&\\
\tilde{h}\in\Omega^{(D-3,1)}_{(2)}({\cal M})&&\end{array}\right.\,.
\ee 
Therefore linearized gravity exhibits a duality symmetry
similar to the electric-magnetic duality of electrodynamics, which
interchanges Bianchi identities and field equations
\cite{Hull:2001}. Tensor gauge fields in
$\Omega^{(D-3,1)}_{(2)}({\cal M})$ have mixed symmetry and were
discussed above in section \ref{mixed}. The right-hand-side of
(\ref{dualcurva}) is represented by \vspace{0.5cm}
\begin{center}\footnotesize
\begin{picture}(0,30)(40,-20)
\multiframe(0,11)(10.5,0){1}(10,10){$\partial$}
\multiframe(10.5,11)(10.5,0){1}(10,10){$\partial$}
\multiframe(0,-0.5)(10.5,0){1}(10,10){}
\multiframe(0,-11)(10.5,0){1}(10,10){}
\put(4.5,-22.5){$\vdots$}\put(4.5,-34){$\vdots$}
\multiframe(0,-45.5)(10.5,0){1}(10,10){}
\multiframe(10.5,-0.5)(10.5,0){1}(10,10){}
\put(30,-10){$\simeq$}
\multiframe(50,-0.5)(10.5,0){1}(10,10){}
\multiframe(50,-11)(10.5,0){1}(10,10){}
\put(54.5,-22.5){$\vdots$}\put(54.5,-34){$\vdots$}
\multiframe(50,-45.5)(10.5,0){1}(10,10){}
\multiframe(50,-56)(10.5,0){1}(10,10){$\partial$}
\multiframe(60.5,-0.5)(10.5,0){1}(10,10){}
\multiframe(60.5,-11)(10.5,0){1}(10,10){$\partial$}
\put(80,-10){.}
\end{picture}\normalsize\end{center}\vspace{1cm}
The appropriate symmetries are automatically implemented by the
antisymmetrizations in (\ref{dualcurva}) since the dual gauge
field $\tilde{h}$ already has the appropriate symmetry
\vspace{0.5cm}
\begin{center}\footnotesize
\begin{picture}(0,25)(40,-20)
\multiframe(0,10)(10.5,0){1}(10,10){}
\multiframe(0,-0.5)(10.5,0){1}(10,10){}
\put(4.5,-12){$\vdots$}\put(4.5,-23.5){$\vdots$}
\multiframe(0,-35)(10.5,0){1}(10,10){}
\multiframe(10.5,10)(10.5,0){1}(10,10){}
\end{picture}\normalsize\,.\end{center}\vspace{0.5cm}
In other words, the two explicit antisymmetrizations in
(\ref{dualcurva}) are sufficient to ensure that the dual tensor
$*R$ possesses the symmetries associated with $Y_{(D-2,2)}^{(2)}$. A
general explanation of this fact will be given at the end of the
next section.

The dual linearized Riemann tensor is invariant under the
transformation \be\delta \tilde{h}=d(S+A)\,\quad\mbox{with}\quad
S\in\Omega^{(D-4,1)}_{(2)}({\cal M})\,,\quad
A\in\Omega^{(D-3,0)}_{(2)}({\cal M})\cong \Omega^{D-3}({\cal
M})\,.\ee The right-hand side of this gauge transformation,
explicitly written in (\ref{gtransfoht}), is represented by
\begin{center}\footnotesize
\begin{picture}(0,37)(40,-20)
\multiframe(0,10)(10.5,0){1}(10,10){}
\multiframe(0,-0.5)(10.5,0){1}(10,10){}
\put(4.5,-12){$\vdots$}\put(4.5,-23.5){$\vdots$}
\multiframe(0,-35)(10.5,0){1}(10,10){$\partial$}
\multiframe(10.5,10)(10.5,0){1}(10,10){}\put(25,7){$\bigoplus$}
\multiframe(40,10)(10.5,0){1}(10,10){}
\multiframe(40,-0.5)(10.5,0){1}(10,10){}
\put(44.5,-12){$\vdots$}\put(44.5,-23.5){$\vdots$}
\multiframe(40,-35)(10.5,0){1}(10,10){}
\multiframe(50.5,10)(10.5,0){1}(10,10){$\partial$}
\end{picture}\normalsize\end{center}\vspace{0.5cm}
In this formalism, the reducibilities (\ref{red1}) and
(\ref{red2}) respectively read (up to coefficient redefinitions)
\ba&\stackrel{(i-1)}{S}=d_L\stackrel{(i)}{S}+d\stackrel{(i)}{A}\,,\quad
\stackrel{(i-1)}{A}=-d_L\stackrel{(i)}{A} \,,\quad (i=2,\ldots, D-2)\,,&\nn\\
&\stackrel{(i)}{S}\in\Omega^{(D-3-i,1)}_{(2)}({\cal M})\,,\quad
\stackrel{(i)}{A}\in\Omega^{D-2-i}({\cal M})\,.&\label{redu}\ea
These reducibilities are a direct consequence of Corollary
\ref{corol}.
%
\subsection{Comparison with electrodynamics}
%
Compared to electromagnetism, linearized gravity presents several
new features. First, there are now two kinds of Bianchi
identities, some of which are algebraic relations (Bianchi I)
while the others are differential equations (Bianchi II). In
electromagnetism, only the latter are present. Second (and perhaps
more importantly), the equation of motion of linearized gravity
theory is an algebraic equation for the curvature (more precisely,
$\mbox{Tr}R=0$). This is natural since the curvature tensor
already contains two derivatives of the gauge field. Moreover, for
higher spin gauge fields $h\in\Omega^{(1,\ldots,1)}_S({\cal M})$
($S\geq 3$) the natural gauge invariant curvature
$d^Sh\in\Omega^{(2,\ldots,2)}_S({\cal M})$ contains $S$
derivatives of the completely symmetric gauge field, hence local
second order equations of motion cannot contain this curvature.
Third, the current conservation in electromagnetism is a direct
consequence of the field equation while for linearized gravity the
Bianchi identities play a crucial role.

In relation to the first remark, the introduction of sources for
linearized gravity seems rather cumbersome to deal with. A natural
proposal is to replace the Bianchi I identities by equations\be
\mbox{Tr}*R=\hat{T}\,,\quad\hat{T}\in\Omega^{D-3,1}_{[2]}({\cal
M})\,.\ee If one uses the terminology of electrodynamics it is
natural to call $\hat{T}$ a ``magnetic" source. If such a dual
source is effectively present, i.e. $\hat{T}\neq 0$, the tensor
$R$ is no longer irreducible under $GL(D,\mathbb R)$, that is to
say $R$ becomes a sum of tensors of different symmetry types and
only one of them has the Riemann tensor symmetries. This seems a
difficult starting point. The linearized Einstein equations read
\be \mbox{Tr}R=\overline{T}\,,\quad
\overline{T}\in\Omega^{1,1}_{[2]}({\cal M})\,.\ee The sources
$\overline{T}$ and $\hat{T}$ respectively couple to the gauge
fields $h$ and $\tilde{h}$. The ``electric" source $\overline{T}$
is a symmetric tensor (related to the energy-momentum tensor) if
the dual source $\hat{T}$ vanishes, since $R\in\Omega^4_2({\cal
M})$ in that case. An other intriguing feature is that a violation
of Bianchi II identities implies a non-conservation of the
linearized energy-momentum tensor because \be \partial^\m
T_{\m\n}=\frac32\,
\partial_{[\m}^{\,}R_{\n\r]}^{\quad\m\r}\,,\ee according to the
linearized Einstein equations (\ref{Einsteinwsource}).

Let us now stress some peculiar features of $D=4$ dimensional
spacetime. From our previous experience with electromagnetism and
our definition of Hodge duality, we naturally expect this
dimension to be privileged. In fact, the analogy between
linearized gravity and electromagnetism is closer in four
dimensions because less independent equations are involved : $*R$
has the same symmetries as the Riemann tensor, thus the dual gauge
field $\tilde{h}$ is a symmetric tensor in $\Omega^2_2({\cal M})$.
So the Hodge duality is a symmetry of the theory only in
four-dimensional spacetime. The dual tensor $*R$ is represented by
a Young diagram of rectangular shape and Proposition
\ref{GenPoinrect} can be used to derive the existence of the dual
potential as a consequence of the field equation $d*R=0$.
%
\section{Generalized Poincar\'e lemma}\label{GenPl}
%
Even if we restrict our attention to completely symmetric tensor
gauge field theories, the Hodge duality operation enforced the use
of the space $\Omega_{(S)}({\cal M})$ of tensors with at most $S$
columns in the previous section. This unavoidable fact requires an
extension of the Proposition \ref{GenPoinrect} to general
irreducible tensors in $\Omega_{(S)}({\cal M})$.
%
\subsection{Generalized cohomology}
%
The {\bf generalized cohomology}\footnote{This definition of
generalized cohomology extends the definition of
``hypercohomology" introduced in \cite{Olver:1983}.} of the
generalized complex $\Omega_{(S)}({\cal M})$ is defined to be the
family of graded vector spaces $H_{(m)}(d)=\oplus_{{\mathbb
Y}^{(S)}} H^{(l_1,\ldots,l_S)}_{(m)}(d)$ with $1\leq m \leq S$
where $H^{(l_1,\ldots,l_S)}_{(m)}(d)$ is the set of $\a\in
\Omega^{(l_1,\ldots,l_S)}_{(S)}({\cal M})$ such that\be d^I\a = 0
\; \; \; \; \forall I \subset \{1,2, \dots, S\} \; \,\vert\, \, \#
I = m\,,\,\,d^I\a\in\Omega_{(S)}({\cal M})\ee with the equivalence
relation \be\a\,\,\,\sim\,\,\,\a\,\,\,+ \sum_{\begin{array}{c}
J \subset \{1,2, \dots, S\}\\
\#J = S - m +1
\end{array}}d^J\b_J\,,\quad\quad \b_J\in
\Omega_{(S)}({\cal M})\,.\label{equivrelatio}\ee Let us stress
that each $\b_J$ is a tensor in an irreducible representation of
$GL(D,\mathbb R)$ such that $d^J\b_J\in
\Omega^{(l_1,\ldots,l_S)}_{(S)}({\cal M})$. In other words, each
irreducible tensor $d^J\b_J$ is represented by a specific diagram
$Y_{(l_1,\ldots,l_S)}^J$ constructed in the following way :
\begin{description}
  \item{1st.} Start from the Young diagram $Y_{(l_1,\ldots,l_S)}^{(S)}$
of the irreducible tensor field
  $\a$.
  \item{2nd.} Remove the lowest cell in $S-m+1$ columns of the
  diagram, making sure that the reminder is still a Young diagram.
  \item{3rd.} Replace all the removed cells with cells
  containing a partial derivative.
\end{description}
The irreducible tensors $\b_J$ are represented by a diagram
obtained at the second step.\\ \noindent 
A less explicit definition of the generalized cohomology is by the
following quotient\be
H_{(m)}(d)=\frac{\bigcap\,\mbox{Ker}\,d^m}{\sum\,\mbox{Im}\,d^{S-m+1}}\,.\ee

We can now state a generalized version of the Poincar\'e lemma,
the proof of which will be postponed to the next subsection
because it is rather lengthy and technical.
\\
\\
\noindent{\bf{Theorem}}
{\it (Generalized Poincar\'e lemma)

Let $Y^{(l_1,\ldots,l_S)}_{(S)}$ be a Young diagram with $l_S\neq 0$ 
and columns
of lengths strictly smaller than $D$ : $l_i<D$, $\forall
i\in\{1,2, \dots, S\}$.
For all $m\in\mathbb N$ such that $1\leq m\leq S$ one has that
\ba
H^{(l_1,\ldots,l_S)}_{(m)}\left(\Omega_{(S)}({\mathbb
R}^D)\right)\cong 0\,.\nn\ea}
\\
\noindent The theorem extends Proposition \ref{GenPoinrect}; the
latter can be recovered retrospectively by the fact that, for
rectangular tensors, there exists only one $d^I\a$ and one $\b_J$.
%
\subsection{Applications to gauge theories}
%
In linearized gravity, one considers the action of nilpotent
operators $d_i$ on the tensors instead of the distinct operators
$d^{\{i\}}$. However, it is possible to show the
useful\begin{proposition}Let $\a$ be an irreducible tensor of
$\Omega_{(S)}({\cal M})$. We have the implication\ba
&&\big(\prod_{ i\in I}d_i\,\,\big)\,\,\a=0\,, \quad\forall
I\subset \{1,2, \dots, S\} \; \,\vert\, \, \# I = m
\nn\\&&\Longrightarrow\quad d^I\a=0\,, \quad\forall I\subset
\{1,2, \dots, S\} \; \,\vert\, \, \# I = m\,.\nn\ea
\label{weaker}\end{proposition}\noindent Therefore, the conditions
appearing in symmetric tensor gauge theories are stronger than the
cocycle condition of $H_{(m)}(d)$ and the coboundary property also
applies.

Now we present the following corollary which is a specific
application of the theorem together with the 
Proposition \ref{weaker}. Its interest resides in its
applicability in linearized gravity field equations (we
anticipated the use of this corollary in the previous subsection).
\begin{corollary}\label{coro}
Let $\kappa\in\Omega_{(S)}({\cal M})$ be an irreducible tensor
field represented by a Young diagram with at least one row of $S$
cells and without any column of length $\geq D-1$. If the tensor
$\kappa$ obeys \ba d_i \kappa=0\,\quad \forall
i\in\{1,\ldots,S\}\,,\nn\ea then
\ba\kappa=\big(\prod\limits^S_{i=1} d_i\big)\lambda\nn\ea where
$\lambda$ belongs to $\Omega_{(S)}({\cal M})$ and the tensor
$\kappa$ is represented by a Young diagram where all the cells of
the first row are filled by partial derivatives.
\end{corollary}

\proof{The essence of the proof is that the two tensors with
diagrams\vspace{0.3cm}
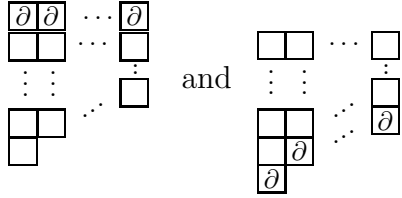
\begin{center}\footnotesize
\begin{picture}(0,37)(0,-24)
\multiframe(-80,11)(10.5,0){1}(10,10){$\partial$}
\multiframe(-80,-0.5)(10.5,0){1}(10,10){}
\put(-75.5,-14){$\vdots$}
\multiframe(-80,-29.5)(10.5,0){1}(10,10){}
\multiframe(-80,-40)(10.5,0){1}(10,10){}
\multiframe(-69.5,11)(10.5,0){1}(10,10){$\partial$}
\multiframe(-69.5,-0.5)(10.5,0){1}(10,10){} \put(-65,-14){$\vdots$}
\multiframe(-69.5,-29.5)(10.5,0){1}(10,10){} 
\put(-54.5,5){$\ldots$}
\put(-52.5,15){$\ldots$}
\multiframe(-38,-0.5)(10.5,0){1}(10,10){}
\multiframe(-38,11)(10.5,0){1}(10,10){$\partial$}
\put(-34,-6){$.$}\put(-34,-3){$.$}
\multiframe(-38,-18)(10.5,0){1}(10,10){}
\put(-50,-20){$.$}\put(-53,-22){$.$}\put(-47,-18){$.$}
\put(-15,-11){\normalsize and}

\multiframe(14,0)(10.5,0){1}(10,10){}
\put(17.5,-13.5){$\vdots$}
\multiframe(14,-29.5)(10.5,0){1}(10,10){}
\multiframe(14,-40)(10.5,0){1}(10,10){}
\multiframe(14,-50.5)(10.5,0){1}(10,10){$\partial$}
\multiframe(24.5,0)(10.5,0){1}(10,10){}
\put(30,-13.5){$\vdots$}
\multiframe(24.5,-29.5)(10.5,0){1}(10,10){}
\multiframe(24.5,-40)(10.5,0){1}(10,10){$\partial$}
\put(40.5,5){$\ldots$}
\multiframe(57,0)(10.5,0){1}(10,10){}\put(61,-6){.}\put(61,-3){.}
\multiframe(57,-18)(10.5,0){1}(10,10){}
\multiframe(57,-28.5)(10.5,0){1}(10,10){$\partial$}
\put(45,-20){$.$}\put(42,-22){$.$}\put(48,-18){$.$}
\put(45,-30){$.$}\put(42,-32){$.$}\put(48,-28){$.$}
\end{picture}\normalsize\end{center}\vspace{.5cm}
are proportional since the initial symmetrization of the partial
derivatives will be cancelled out by the
antisymmetrization in the columns that immediately follows for two
attached columns of different length (if they have the same length
the partial derivatives are in the same row and the symmetrization
is automatic). By induction, starting from the left, one proves
that this is true for an arbitrary number of columns. This
argument remains true if we add smaller columns on the right of
the Young diagram.

The last subtlety in the corollary is that antisymmetrization in
each column ($\kappa=(\prod d_i)\lambda$) automatically provides the
appropriate Young symmetrization since $\lambda$ has the
appropriate symmetry. This can be easily checked by performing a
complete antisymmetrization of the tensor $\kappa$ in the entries
of a column and another entry which is on its right. The result
automatically vanishes because the index in the column at the
right is either
\begin{itemize}
\item attached to a partial derivative, in which case the antisymmetrization
contains two
partial derivatives, or
\item attached to the tensor $\lambda$.
In this case, antisymmetrization over the indices of the column
except the one in the first row (corresponding to a partial
derivative) causes an antisymmetrization of the
tensor $\lambda$ in the entries of a column and another entry
which is on the right-hand side. The result vanishes since $\l$
has the symmetry properties corresponding to the diagram obtained
after eliminating the first row of $\kappa$.
\end{itemize}
}

This last discussion can be summarized by the operator formula
\be
\prod\limits^S_{i=1}
d_i\circ\mathbf{Y}_{(l_1,\ldots,l_S)}^{(S)}\,
\propto\,
\mathbf{Y}_{(l_1+1,\ldots,l_S+1)}^{(S)} \circ\partial^S
\circ\mathbf{Y}_{(l_1,\ldots,l_S)}^{(S)}\,,
\ee 
where
$\partial^S$ are $S$ partial derivatives with indices
corresponding to the first row of a given Young diagram.

We now present another corollary, which determines the
reducibility identities for the mixed symmetry type gauge field.
\begin{corollary}\label{corol}
Let $\l\in\Omega_{(2)}({\cal M})$ be a sum of two irreducible
tensors \\ \noindent $\l_1\in\Omega^{(l_1-1,l_2)}_{(2)}({\cal M})$ and
$\l_2\in\Omega^{(l_1,l_2-1)}_{(2)}({\cal M})$ with $l_1\geq l_2$
($\l_1=0$ if $l_1=l_2$). Then, \ba
\sum\limits^2_{i=1}d^{\{i\}}\l_i=0\,\quad\Rightarrow\quad
\l_i\,=\,\sum\limits_{j=1}^2d^{\{j\}} \m_{ij} \quad
(i=1,2)\,,\nn\ea where
\begin{description}
  \item[-] $\m_{11}\in\Omega^{(l_1-2,l_2)}_{(2)}({\cal M})$ (which vanishes
if $l_1\leq l_2+1$),
  \item[-] $\m_{12},\m_{2,1}\in\Omega^{(l_1-1,l_2-1)}_{(2)}({\cal M})$
  ($\m_{1,2}=0$ if $l_1= l_2$), and
  \item[-] $\m_{22}\in\Omega^{(l_1,l_2-2)}_{(2)}({\cal M})$.
\end{description}
Furthermore, if $l_1> l_2$ we can assume, without loss of
generality, that \ba\m_{21}=-\m_{12}\,.\nn\ea
\end{corollary}
\proof{We apply $d^{\{1\}}$ and $d^{\{2\}}$ to the equation
$\sum\limits^2_{i=1}d^{\{i\}}\l_i=0$ and obtain
$0=d^{1,2}\l_i\propto d_1 d_2 \l_i$ in view of the remarks
following Corollary \ref{coro}. From the theorem, we deduce that
$\l_i\,=\,\sum\limits_{j=1}^2d^{\{j\}} \m_{ij} $ with tensors
$\m_{ij}$ in the appropriate spaces given in Corollary
\ref{corol}. The fact they vanish agrees with the rule given
above.

To finish the proof we should consider the case $l_1>l_2$.
Assembling the results together,
$\sum\limits^2_{i=1}d^{\{i\}}\l_i=d^{\{1,2\}} (\m_{12}+\m_{21})=0$
due to Proposition \ref{commutprop}. Thus, $d_1 d_2
(\m_{12}+\m_{21})=0$. Using Corollary \ref{corol} again, one
obtains $\m_{12}+\m_{21}=\sum\limits^2_{k=1}d^{\{k\}}\n_k$ with
$\n_1\in\Omega^{(l_1-2,l_2-1)}_{(2)}({\cal M})$ and
$\n_2\in\Omega^{(l_1-1,l_2-2)}_{(2)}({\cal M})$ ($\n_1=0$ if
$l_1=l_2$). Hence we can make the redefinitions
$\m_{12}\rightarrow\m'_{12}=\m_{12}-d^{\{2\}}\n_2$ and
$\m_{21}\rightarrow\m'_{12}=\m_{21}-d^{\{1\}}\n_1$ which do not
affect $\l$, in such a way that we have $\m'_{21}=-\m'_{12}$.}
This proposition can be generalized to give a full proof of the
gauge reducibility rules given in \cite{Labastida:1986} and will
be reviewed in subsection \ref{reduc}.
%
\section{Arbitrary Young symmetry type gauge field theories}
\label{arbitraryYoung}
%
We now generalize the results of section \ref{fieldequs} to
arbitrary irreducible tensor representations of $GL(D,\mathbb R)$.
The discussion presented below fits into the approach followed by
\cite{Aulakh:1986} for two columns ($S=2$) and by
\cite{Labastida:1986,Hull:2001} for an arbitrary number of
columns. The interest of this section lies in the translation of
these old results in the present mathematical language and in the
use of the generalized Poincar\'e lemma for a more systematic
mathematical foundation.
%
\subsection{Bianchi identities}
%
Firstly, we generalize our previous discussion on linearized
gravity by introducing a tensor $K$, which is the future
curvature. A priori, $K$ is a multiform of
$\Omega^{l_1,\ldots,l_S}_{[S]}({\cal M})$ ($l_S\neq 0$) with
$1\leq l_j\leq l_i<D$ for $i\leq j$. Secondly, we suppose the
(algebraic) {\bf Bianchi I} relations to be
\be\mbox{Tr}_{ij}\{\,*_i\,K\}=0\,,\quad \forall i,j:\,\, 1\leq
i<j\leq S\,,\label{BianchI}\ee in order to obtain, from
Proposition \ref{Schurm}, that $K$ is an irreducible tensor under
$GL(D,\mathbb R)$ belonging to
$\Omega^{(l_1,\ldots,l_S)}_{(S)}({\cal M})$. Thirdly, we define
the (differential) {\bf Bianchi II} relations as \be
d_iK=0\,,\quad\forall i:\,\, 1\leq i\leq S\,,\label{BianchII}\ee
in such a way that, from Corollary \ref{coro}, one obtains\be
K=d_1d_2\ldots d_S \k\,.\ee In this case, the curvature is indeed
a natural object for describing a theory with gauge fields $\k\in
\Omega^{(l_1-1,\ldots,l_S-1)}_{(S)}({\cal M})$. The gauge
invariances are then \be \k\rightarrow
\k+d^{\{i\}}\b_i\,,\label{gtransfo}\ee where the gauge parameters
$\b_i$ are irreducible tensors $\b_i$ in \\ \noindent
$\Omega^{(l_1-1,\ldots,l_i-2,\ldots,l_S-1)}_{(S)}({\cal M})$ for
any $i$ such that $l_i\geq 2$ (and $l_i>l_{i-1}$), as follows from
our theorem and Proposition \ref{weaker}.
%
\subsection{Reducibilities}\label{reduc}
%

The gauge transformations (\ref{gtransfo}) are generally
reducible, i.e. $d^{\{j\}}\b_j\equiv 0$ for non-vanishing
irreducible tensors $\b_j\neq 0$. The procedure followed in the
proof of Corollary \ref{corol} can be applied to the general case.
This generates the inductive rules of \cite{Labastida:1986} to
form the $(i+1)$-th generation reducibility parameters
$\stackrel{(i+1)}{\b}_{j_1j_2\ldots j_{i+1}}$ from the $i$-th
generation paramaters $\stackrel{(i)}{\b}_{j_1j_2\ldots j_i}$:

\begin{itemize}

\item $\underline{i=1}$

\begin{description}
  \item[(A)] Start with the Young diagram $Y_{(l_1-1,\ldots,l_S-1)}^{(S)}$
 corresponding to the tensor gauge field $\kappa$.
  \item[(B)] Remove a box from a row such
that the result is a standard Young diagram. In other words, the
gauge parameters are taken to be the first reducibility
parameters: $\b_j=\stackrel{(1)}{\b}_j$.
\end{description}

\item $\underline{i\rightarrow i+1}$

\begin{description}
  \item[(C)] Remove a box from a row which has not previously had a box removed
(in forming the lower generations of reducibility parameters) such
that the result is a standard Young diagram.
  \item[(D)] There is one and only one
reducibility parameter for each Young diagram.
\end{description}

\end{itemize}

The Labastida-Morris rules (A)-(D) provide the complete BRST
spectrum with the full tower of ghosts of ghosts.\\ \noindent
More explicitly, the chain of reducibilities is \be
\stackrel{(i)}{\b}_{j_1j_2\ldots
j_i}=d^{\{j_{i+1}\}}\stackrel{(i+1)}{\b}_{j_1j_2\ldots
j_ij_{i+1}}=0\,,\quad (i=1,2,\ldots,r)\ee where $r=l_1-1$ is the
number of rows of $\k$. The chain is of length $r$ because at each
step one removes a box from a row which has not been chosen
before. We can see that the order of reducibility of the gauge
transformations (\ref{gtransfo}) is equal to $l_1-2$. For $S=1$,
we recover the fact that a $p$-form gauge field theory ($l_1=p+1$)
has its order of reducibility equal to $p-1$.

The subscripts of the $i$-th reducibility parameter
$\stackrel{(i)}{\b}_{j_1j_2\ldots j_i}$ belong to the set
$\{1,\ldots,S\}$. These determine the Young diagram corresponding
to the irreducible tensor $\stackrel{(i)}{\b}_{j_1j_2\ldots j_i}$:
reading from the left to the right, the subscripts give the
successive columns from which to remove the bottom box following
the rules (A)-(C). A reducibility parameter
$\stackrel{(i)}{\b}_{j_1j_2\ldots j_i}$ vanishes if these rules
are not fulfilled. Furthermore, they are antisymmetric for any
pair of different indices\be \stackrel{(i)}{\b}_{j_1\ldots
j_k\ldots j_l\ldots j_i}=-\stackrel{(i)}{\b}_{j_1\ldots j_l\ldots
j_k\ldots j_i}\,,\quad \forall j_l\neq j_k\,.\label{minusign}\ee
This property ensures the rule (D) and provides the correct signs
to fulfill the reducibilities. Indeed, \be
d^{\{j_i\}}\stackrel{(i)}{\b}_{j_1j_2\ldots
j_i}=d^{\{j_i\}}d^{\{j_{i+1}\}}\stackrel{(i+1)}{\b}_{j_1j_2\ldots
j_ij_{i+1}}=0\,,\ee due to Proposition \ref{commutprop} and
equation (\ref{minusign}).
%
\subsection{Field equations and dualisation properties}
%
We make the important following assumption concerning the positive integers 
$l_i$ $(i=1,\ldots,S)$ associated to $K\in\O_{[S]}^{l_1,\ldots,l_S}$ : 
\be 
l_i+l_j\leq D\,,\quad
\forall\, i,j\label{hypoth} 
\ee 
and take the field equations to be
in that case \be \mbox{Tr}_{ij}\{K\}=0\,,\quad \forall
\,i,j\,.\label{FIeld} \ee Indeed, if $l_1+l_2> D$ and if the
equation (\ref{FIeld}) holds, then the curvature $K$ identically
vanishes, as is well known when studying irreps of $O(D-1,1)$.
This property is a particular instance of Proposition
\ref{tracepower} (for $m=0$).

The field equations (\ref{FIeld}) combined with the Bianchi I
identities (\ref{BianchI}) state that the curvature $K$ is a
tensor irreducible under $O(D-1,1)$.

To any non-empty subset $I\subset\{1,2, \dots, S\}$ ($\#I=m$), we
associate a Hodge duality operator\be *_I\equiv  \prod_{k \in I}
*_k \,.\ee The dual $*_I K$ of the curvature is a multiform in
$\Omega^{\ell_1,\ldots,\ell_S}_{[S]}({\cal M})$, where the
lengths $\ell_i$ are defined in equation (\ref{ell}).

The Bianchi I identities (\ref{BianchI}) together with the field
equations (\ref{FIeld}) imply the relations 
\be
\mbox{Tr}_{ij}\,\{*_i\,(*_I K\,)\}\,=\,0\,,\quad \forall i,j:
\ell_j\leq\ell_i
\label{from}
\ee  
where $\ell_i$ is the length (\ref{ell}) of the i-th column of the dual 
tensor $*_I K$. 
Indeed, let be $i$ and $j$ such that $\ell_j\leq\ell_i$. 
There are essentially four possibilities:
\begin{itemize}
  \item $i\not\in I$ and
    \begin{itemize}
    \item $\underline{j\not\in I}$: Then $l_j\leq l_i$ and the 
    Bianchi I identities (\ref{BianchI}) 
    are equivalent to (\ref{from}) since $\mbox{Tr}_{ij}$ and $*_k$
    commute if $i\neq k$ and $j\neq k$.
    \item $\underline{j\in I}$: Then one should have $D-l_j\leq l_i$
    which means that $D\leq l_i+l_j$, in contradiction
    with the hypothesis (\ref{hypoth}) except for the case where 
    $l_i+l_j=D$. From Proposition
\ref{tracepower}, we deduce that in such a case the field equations
(\ref{FIeld}) are equivalent to (\ref{from}).
\end{itemize}
  \item $i\in I$ and
    \begin{itemize}
    \item $\underline{j\not\in I}$: We have $l_j\leq D-l_i$ which is
     equivalent to $l_i+l_j\leq D$. 
    The field equations
    (\ref{FIeld}) are of course equivalent to (\ref{from}) 
    since $*_i^2 K=\pm K$.
    \item $\underline{j\in I}$: We have $D-l_j\leq D-l_i$ which is
    equivalent to $l_i\leq l_j$. The Bianchi I identities 
    $\mbox{Tr}_{j\,i}\{*_j K\}=0$ are therefore satisfied and equivalent 
    to (\ref{from}) 
    because $\mbox{Tr}_{ij}=\mbox{Tr}_{ji}$.
    \end{itemize}
\end{itemize}

Let $\widetilde{Y}_{(\l_1,\ldots,\l_S)}^I$ be the Young diagram
dual to $Y_{(l_1,\ldots,l_S)}^{(S)}$. We define
$\widetilde{K}_I$ to be the multiform in
$\Omega^{\l_1,\ldots,\l_S}_{[S]}({\cal M})$ obtained after reordering the
 columns of $*_I K$. 
The identity (\ref{from}) can then be formulated as
\be
{\mbox{Tr}}_{ij}\{*_i \widetilde{K}_I\}=0\,,~\forall ~i,j~:~1\leq i<j\leq S\,.
\label{frombis}
\ee
Due to Proposition
\ref{Schurm}, it follows from (\ref{frombis}) that the tensor 
$\widetilde{K}_I$ is irreducible under $GL(D,\mathbb R)$
\be 
\widetilde{K}_I\,\in\,
\Omega^{(\l_1,\ldots,\l_S)}_{(S)}({\cal M})\,.
\label{inom}
\ee

Now we use the property (\ref{makeuse}) to deduce from the Bianchi
II identities (\ref{BianchII}) and the field equations
(\ref{FIeld}) that $d_i(*_iK)=0$ for any $i$. 
Therefore 
\be
d_i(*_I{K})=0\,,\quad \forall i\in\{1,\ldots,S\}\,,
\label{folows}
\ee 
because $d_i$ and $*_j$
commute if $i\neq j$, and either
\begin{itemize}
  \item $i\not\in I$ so (\ref{folows}) follows from $d_iK=0$,
  or
  \item $i\in I$ and then (\ref{folows}) is a consequence of $d_i*_iK=0$.
\end{itemize}

In other words, any dual tensor $\widetilde{K}_I$ satisfies
(on-shell) its own Bianchi II identity (\ref{folows}) which,
together with (\ref{inom}), implies the (local) existence of a
dual gauge field $\tilde{\k}_I$ such that the Hodge dual of the
curvature is itself a curvature
\be 
\widetilde{K}_I=d_1d_2\ldots
d_S \tilde{\k}_I
\ee for some gauge field \be\tilde{\k}_I\in
\Omega^{(\l_1-1,\ldots,\l_S-1)}_{\quad(S)}({\cal M}) \,.\ee The
Hodge operators $*_I$ therefore relate different free field
theories of arbitrary tensor gauge fields, extending the
electric-magnetic duality property of electrodynamics.

In the same way, we obtain the field equations of the dual theory
\be 
\mbox{Tr}^{m_{ij}}_{ij}\,\{\,*_I K\,\}\,=\,0\,,\quad \forall ~
i,j:\,i<j\,
\label{genfieldequ}
\ee 
where 
\be 
m_{ij}\equiv
\left\{\begin{array}{lll}1+D-l_i-l_j\quad&\mbox{if}\,\,i\,\,\mbox{and}\,\,j\in
I\,,&\\
1\,\quad&\mbox{if}\,\,i\,\,\mbox{or}\,\,j\not\in
I\,.&\end{array}\right.
\ee 
Indeed, since the trace is symmetric in
$i$ and $j$ we must consider only three distinct cases:
\begin{itemize}
  \item $\underline{i\not\in I}$ and $\underline{j\not\in I}$: 
The starting field
equation (\ref{FIeld}) is naturally equivalent to the dual field
equation (\ref{genfieldequ}).
  \item $\underline{i\in I}$ and
    \begin{itemize}
    \item $\underline{j\not\in I}$: If $i<j$ the Bianchi I
    relation (\ref{BianchI}) is satisfied and it implies (\ref{genfieldequ}).
    \item $\underline{j\in I}$: A direct use of Proposition
    \ref{tracepower} leads from the field equation
    (\ref{FIeld}) to (\ref{genfieldequ}).
\end{itemize}
\end{itemize}
We can summarize the algebraic part of the previous discussions in
terms of a remark on tensorial irreps of $0(D-1,1)$.\\ \noindent
{\bf Remark} : {\it Let $I\subset \{1,\ldots,S\}$ be a non-empty subset. Let
$\widetilde{Y}_{(\l_1,\ldots,\l_S)}^I$ be the Young diagram dual
to $Y_{(l_1,\ldots,l_S)}^{(S)}$.
If $\a\in\Omega^{(l_1,\ldots,l_S)}_{(S)}({\cal M})$ is a tensor in
the irreducible representation of $O(D-1,1)$ associated to the
Young diagram $Y_{(l_1,\ldots,l_S)}^{(S)}$, then the dual tensor
$\widetilde{\a}_I\in\Omega^{(\l_1,\ldots,\l_S)}_{(S)}({\cal M})$
is in the irreducible representation of $O(D-1,1)$ associated to
the Young diagram $\widetilde{Y}_{(\l_1,\ldots,\l_S)}^I$.}

As one can see, a seemingly odd feature of some dual field
theories is that their field equations (\ref{genfieldequ}) are not
of the same type as (\ref{FIeld}). In fact, the dual field
equations are of the type (\ref{FIeld}) for all $I$ only in the
exceptional case where $D$ is even and $l_i=l_j=D/2$. Note that
this condition is satisfied for free gauge theories of completely
symmetric tensors in $D=4$ flat space. The point is that
$\ell_i+\ell_j=2D-l_i-l_j\geq D$ for $i,j\in I$, therefore the
property (\ref{hypoth}) is generally not satisfied by the dual
tensor $\widetilde{K}_I$.

To end up, we generalize the field equation (\ref{FIeld}) to the
case where the hypothesis (\ref{hypoth}) is not satisfied. A
natural idea is that when $l_i+l_j> D$ for a curvature tensor
$K\in\Omega^{(l_1,\ldots,l_S)}_{(S)}({\cal M})$ ($l_S\neq 0$), 
the corresponding fields equations are \cite{Hull:2001}\be
\mbox{Tr}^{1+l_i+l_j-D}_{ij}\,\{\,K\,\}\,=\,0\,.\ee
%

\section*{Acknowledgements}
%
We are grateful to M. Henneaux for a crucial remark at the initial
stage of this work. We would like to thank M. Henneaux and C.
Schomblond for their careful reading and comments on intermediate
versions of this paper. X.B. acknowledges D. Tonei for her help in
english and the organizers of the conference ``Rencontres
Math\'ematiques de Glanon (6\`eme \'edition)" where this work was 
presented \cite{Glanon}.

This work was supported in part by the ``Actions de Recherche
Concert{\'e}es" of the ``Direction de la Recherche Scientifique -
Communaut\'e Francaise de Belgique", IISN-Belgium (convention
4.4505.86), a ``P\^ole d'Attraction Interuniversitaire" (Belgium)
and the European Commission RTN programme HPRN-CT-2000-00131, in
which we are associated with K.U. Leuven.
%
\appendix
\label{appendice}
%
\section{Inductive proof of the generalized Poincar\'e lemma}
\label{InduGeneralPoin}
%
The proof of the generalized Poincar\'e
lemma that we give hereafter is inductive in several directions.
The first induction parameter is the number $S$ of columns;     
in section \ref{subA1} we start from the standard Poincar\'e lemma, 
{\it i.e.} $S=1$, and compute the generalized cohomologies when a cell
is added in a new, second column, {\it i.e.} $S=2$. 
The second induction parameter is the number $\ell$ of cells in the new
(second) column. Thus section \ref{subA1}, which gives a (pictorial) 
proof that 
the cohomological groups $\Omega^{(*,1)}_{(2)}({\mathbb R}^D)$ are trivial, 
also provides the starting point for the induction on $\ell$, 
keeping $S=2$ fixed. 

The inductive proof of the vanishing of 
$H_{(*)}^{(*,*)}\left(\Omega_{(2)}({\mathbb R}^D)\right)$ 
 is then given in sections \ref{subA2}. 
This proof of  
$H_{(*)}^{(*,*)}\left(\Omega_{(2)}({\mathbb R}^D)\right)\cong 0$ is purely 
algebraic and does not contain any  pictorial description, this time.
However, for a better understanding of the algebraic demonstration, 
a pictorial translation of most of the results obtained in  \ref{subA2} is  
furnished in section \ref{subA3}.

The inductive progression we have sketched above is the one used to obtain 
the proof of the vanishing
$H^{(*,\ldots,*)}_{(*)}(\Omega_{(*)}({\mathbb R}^D))\cong 0$, 
for diagrams obeying the assumptions of the Theorem (section \ref{GenPl}).
This time, instead of progressing from $S=1$ to $S=2$ and then from a 
length-$(\ell-1)$ to a length-$\ell$ second column, 
we go from $S$ to $S+1$ columns and subsequently, keeping
the number of columns fixed to $S+1$, we increase the length of the last 
$(S+1)$-th column.  

Since this progression, exposed in detail in sections \ref{subA1}, 
\ref{subA2} and \ref{subA3}, provides the proof of our generalized  
Poincar\'e lemma, we only summarize those results in section \ref{section3} 
and cast our Theorem in precise mathematical terms.
%
\subsection{Generalized cohomology in $\Omega^{(*,1)}_{(2)}({\mathbb R}^D)$}
\label{subA1}
%
Using the standard Poincar\'e lemma ($S=1$), we begin by providing a 
pictorial proof that the two cohomologies
$H_{(1)}^{(n,1)}\left(\Omega_{(2)}({\mathbb R}^D)\right)$ and
$H_{(2)}^{(n,1)}\left(\Omega_{(2)}({\mathbb R}^D)\right)$ are
trivial for $0<n< D$, {\it{i.e.}} that 
\begin{itemize}
\item[$\bullet$]
(1) 
\be d^{\{i\}}~
\begin{picture}(30,30)(0,0)
\multiframe(0,10)(10.5,0){1}(10,10){\ft$1$}
\multiframe(10.5,10)(10.5,0){1}(17,10){\ft$n\!\!+\!\!1$}
\multiframe(0,-0.5)(10.5,0){1}(10,10){\ft$2$}
\multiframe(0,-18)(10.5,0){1}(10,17){$ $} \put(4,-13.5){$\vdots$}
\multiframe(0,-28.5)(10.5,0){1}(10,10){\ft$n$}
\end{picture}
\vspace{1cm} ~=~0~\, , ~~~~~~~~i=1,2 
\label{cond1} 
\ee 
implies
\be
\vspace{1cm}  
\begin{picture}(30,30)(0,0)
\multiframe(0,10)(10.5,0){1}(10,10){\ft$1$}
\multiframe(10.5,10)(10.5,0){1}(17,10){\ft$n\!\!+\!\!1$}
\multiframe(0,-0.5)(10.5,0){1}(10,10){\ft$2$}
\multiframe(0,-18)(10.5,0){1}(10,17){$ $} \put(4,-13.5){$\vdots$}
\multiframe(0,-28.5)(10.5,0){1}(10,10){\ft$n$}
\end{picture}
~=~~ 
\begin{picture}(30,30)(0,0)
\multiframe(0,10)(10.5,0){1}(10,10){\ft$1$}
\multiframe(10.5,10)(10.5,0){1}(10,10){\ft$\partial$}
\multiframe(0,-0.5)(10.5,0){1}(10,10){\ft$2$}
\multiframe(0,-18)(10.5,0){1}(10,17){$ $} \put(4,-13.5){$\vdots$}
\multiframe(0,-28.5)(10.5,0){1}(10,10){\ft$\partial$}
\end{picture}
\label{resu1} \ee 
and 
\item[$\bullet$]
(2) 
\be d^{\{1,2\}}~ 
\begin{picture}(25,30)(0,0)
\multiframe(0,10)(10.5,0){1}(10,10){\ft$1$}
\multiframe(10.5,10)(10.5,0){1}(17,10){\ft$n\!\!+\!\!1$}
\multiframe(0,-0.5)(10.5,0){1}(10,10){\ft$2$}
\multiframe(0,-18)(10.5,0){1}(10,17){$ $} \put(4,-13.5){$\vdots$}
\multiframe(0,-28.5)(10.5,0){1}(10,10){\ft$n$}
\end{picture}\vspace{1cm}
~~=~0\, 
\label{cond2} 
\ee 
implies 
\be
\begin{picture}(25,30)(0,0)
\multiframe(0,10)(10.5,0){1}(10,10){\ft$1$}
\multiframe(10.5,10)(10.5,0){1}(17,10){\ft$n\!\!+\!\!1$}
\multiframe(0,-0.5)(10.5,0){1}(10,10){\ft$2$}
\multiframe(0,-18)(10.5,0){1}(10,17){$ $} \put(4,-13.5){$\vdots$}
\multiframe(0,-28.5)(10.5,0){1}(10,10){\ft$n$}
\end{picture}\vspace{1cm}
~=~~ 
\begin{picture}(25,30)(0,0)
\multiframe(0,10)(10.5,0){1}(10,10){\ft$1$}
\multiframe(10.5,10)(10.5,0){1}(10,10){\ft$\partial$}
\multiframe(0,-0.5)(10.5,0){1}(10,10){\ft$2$}
\multiframe(0,-18)(10.5,0){1}(10,17){$ $} \put(4,-13.5){$\vdots$}
\multiframe(0,-28.5)(10.5,0){1}(10,10){\ft$n$}
\end{picture}+~~
\begin{picture}(25,30)(0,0)
\multiframe(0,10)(10.5,0){1}(10,10){\ft$1$}
\multiframe(10.5,10)(10.5,0){1}(17,10){\ft$n\!\!+\!\!1$}
\multiframe(0,-0.5)(10.5,0){1}(10,10){\ft$2$}
\multiframe(0,-18)(10.5,0){1}(10,17){$ $} \put(4,-13.5){$\vdots$}
\multiframe(0,-28.5)(10.5,0){1}(10,10){\ft$\partial$}
\end{picture}~~. \vspace{.5cm}
\label{resu2} \ee 
\end{itemize}
The numbers in the cells are irrelevant, they
simply signal the length of the columns. For clarity we recall the
following convention that, whenever a Young tableau $Y$ appears
with certain boxes filled in with partial derivatives $\partial$,
one takes a field with the representation of the Young tableau
obtained by removing all the $\partial$-boxes from $Y$. One
differentiates this new field as many times as there are
derivatives in $Y$ and then project the result on the Young
symmetry of $Y$.
%
\subsubsection{First cohomology group}
%
For the two different possible values of $i$ in (\ref{cond1}) we
have the following two conditions on the field $(n,1)$ :
\begin{itemize}
\item
\hspace*{2mm} 
\begin{picture}(25,30)(0,0)
\multiframe(0,10)(10.5,0){1}(10,10){\ft$1$}
\multiframe(10.5,10)(10.5,0){1}(17,10){\ft$n\!\!+\!\!1$}
\multiframe(0,-7.5)(10.5,0){1}(10,17){$ $} \put(3.5,-4){$\vdots$}
\multiframe(0,-18)(10.5,0){1}(10,10){\ft$n$}
\multiframe(0,-28.5)(10.5,0){1}(10,10){\ft$\partial$}
\end{picture}
\vspace{1cm}
~~~$=0$~~~ for $i=1$
\\
and
\item
\hspace*{2mm}  
\begin{picture}(25,30)(0,0)
\multiframe(0,10)(10.5,0){1}(10,10){\ft$1$}
\multiframe(10.5,10)(10.5,0){1}(17,10){\ft$n\!\!+\!\!1$}
\multiframe(10.5,-0.5)(10.5,0){1}(17,10){\ft$\partial$}
\multiframe(0,-7.5)(10.5,0){1}(10,17){$ $} \put(3.5,-4){$\vdots$}
\multiframe(0,-18)(10.5,0){1}(10,10){\ft$n$}
\end{picture}
\vspace{1cm}
~~~$=0$~~~ for $i=2$~.
\end{itemize}

\noindent
The first condition is treated now : one considers the index of
the second column as a spectator, which yields 
\ba 
\begin{picture}(25,30)(0,0)
\multiframe(0,10)(10.5,0){1}(17,10){\ft$1$}
\multiframe(0,-0.5)(10.5,0){1}(17,10){\ft$2$}
\multiframe(17.5,10)(10.5,0){1}(17,10){\ft$n\!\!+\!\!1$}
\multiframe(0,-17)(10.5,0){1}(17,16){$ $} \put(7,-13){$\vdots$}
\multiframe(0,-27.5)(10.5,0){1}(17,10){\ft$n$}
\multiframe(0,-38)(10.5,0){1}(17,10){\ft$\partial$}
\end{picture}
~~~~\simeq~~~~ 
\begin{picture}(25,30)(0,0)
\multiframe(0,10)(10.5,0){1}(17,10){\ft$1$}
\multiframe(0,-0.5)(10.5,0){1}(17,10){\ft$2$}
\multiframe(0,-17)(10.5,0){1}(17,16){$ $} \put(7,-13){$\vdots$}
\multiframe(0,-27.5)(10.5,0){1}(17,10){\ft$n$}
\multiframe(0,-38)(10.5,0){1}(17,10){\ft$\partial$}
\end{picture}
\otimes ~~  
\begin{picture}(15,15)(0,0)
\multiframe(0,-0.5)(10.5,0){1}(17,10){\ft$n\!\!+\!\!1$}
\end{picture}\quad=\,0\nn
\ea
\\
\\
\\
where the symbol $\simeq$ means that there is an implicit
projection using ${\bf{Y}}$ on the right-hand side in order to
agree with the left-hand side (in other words the symbol $\simeq$
replaces the expression $=\mathbf Y$). The Poincar\'e Lemma is
used for the first column to write, using branching rules for
$GL(D,\mathbb R)$ :  
\begin{center}
\begin{picture}(25,60)(0,-30)
\multiframe(-130,10)(10.5,0){1}(10,10){\ft$1$}
\multiframe(-119.5,10)(10.5,0){1}(17,10){\ft$n\!\!+\!\!1$}
\multiframe(-130,-7.5)(10.5,0){1}(10,17){$ $} 
\multiframe(-130,-18)(10.5,0){1}(10,10){\ft$n$}\put(-127,-4){$\vdots$}
\put(-95,0){$\simeq$} 
\multiframe(-80,10)(10.5,0){1}(17,10){\ft$1$}
\multiframe(-80,-0.5)(10.5,0){1}(17,10){\ft$2$}
\multiframe(-80,-17)(10.5,0){1}(17,16){$ $} \put(-73,-13){$\vdots$}
\multiframe(-80,-27.5)(10.5,0){1}(17,10){\ft$n\!\!-\!\!1$}
\multiframe(-80,-38)(10.5,0){1}(17,10){\ft$\partial$}
\put(-55,0){$\otimes$} 
\multiframe(-40,-0.5)(10.5,0){1}(10,10){\ft$n$}
\put(-25,0){$\simeq$}
\multiframe(-10,0)(10.5,0){1}(10,10){\ft$\partial$}
\put(5,0){$\otimes~ \Big($} 
\multiframe(25,10)(10.5,0){1}(17,10){\ft$1$}
\multiframe(25,-0.5)(10.5,0){1}(17,10){\ft$2$}
\multiframe(25,-17)(10.5,0){1}(17,16){$ $} \put(32,-13){$\vdots$}
\multiframe(25,-27.5)(10.5,0){1}(17,10){\ft$n\!\!-\!\!1$}
\put(50,0){$\otimes$}
\multiframe(65,-0.5)(10.5,0){1}(10,10){\ft$n$}
\put(80,0){$\Big)~~.$}
\end{picture}
\end{center}
In the last step, we have undone the manifest antisymmetrization
with the index carrying the partial derivative; we are more
interested in the symmetries of the tensor under the derivative.

We first perform the product in the brackets to obtain a sum of
different types of irreducible tensors. Then, we perform the
product with the partial derivative to get 
\ba 
\begin{picture}(25,30)(0,0)
\multiframe(0,10)(10.5,0){1}(17,10){\ft$1$}
\multiframe(17.25,10)(10.5,0){1}(17,10){\ft$n\!\!+\!\!1$}
\multiframe(0,-0.5)(10.5,0){1}(17,10){\ft$2$}
\multiframe(0,-17)(10.5,0){1}(17,16){$ $} \put(7,-13.5){$\vdots$}
\multiframe(0,-27.5)(10.5,0){1}(17,10){\ft$n\!\!-\!\!1$}
\multiframe(0,-38)(10.5,0){1}(17,10){\ft$n$}
\end{picture}
\vspace{2cm} ~~~~\simeq~~~~ 
\begin{picture}(30,30)(0,0)
\multiframe(0,10)(10.5,0){1}(17,10){\ft$1$}
\multiframe(17.25,10)(10.5,0){1}(17,10){\ft$\partial$}
\multiframe(0,-0.5)(10.5,0){1}(17,10){\ft$2$}
\multiframe(0,-17)(10.5,0){1}(17,16){$ $} \put(7,-13){$\vdots$}
\multiframe(0,-27.5)(10.5,0){1}(17,10){\ft$n\!\!-\!\!1$}
\multiframe(0,-38)(10.5,0){1}(17,10){\ft$n$}
\end{picture}
\hspace*{2mm}\oplus\hspace*{2mm} 
\begin{picture}(30,30)(0,0)
\multiframe(0,10)(10.5,0){1}(17,10){\ft$1$}
\multiframe(17.25,10)(10.5,0){1}(17,10){\ft$n$}
\multiframe(0,-0.5)(10.5,0){1}(17,10){\ft$2$}
\multiframe(0,-17)(10.5,0){1}(17,16){$ $} \put(7,-13){$\vdots$}
\multiframe(0,-27.5)(10.5,0){1}(17,10){\ft$n\!\!-\!\!1$}
\multiframe(0,-38)(10.5,0){1}(17,10){\ft$\partial$}
\end{picture}
\hspace*{2mm}\oplus\hspace*{2mm} 
\begin{picture}(30,30)(0,0)
\multiframe(0,10)(10.5,0){1}(17,10){\ft$1$}
\multiframe(0,-0.5)(10.5,0){1}(17,10){\ft$2$}
\multiframe(0,-17)(10.5,0){1}(17,26){$ $} \put(7,-13){$\vdots$}
\multiframe(0,-27.5)(10.5,0){1}(17,10){\ft$n\!\!-\!\!1$}
\multiframe(0,-38)(10.5,0){1}(17,10){\ft$n$}
\multiframe(0,-48.5)(10.5,0){1}(17,10){\ft$\partial$}
\end{picture}\,.
\label{decom1} \ea
\\
\\
\\
\noindent The last term in the above equation (\ref{decom1}) does
not match the symmetry of the left-hand-side, so it must vanish.
Using the Poincar\'e lemma, which is applicable since one is not
in top form degree : $n<D$, one obtains 
\be 
\begin{picture}(25,30)(0,0)
\multiframe(0,20)(10.5,0){1}(10,10){\ft$1$}
\multiframe(0,9.5)(10.5,0){1}(10,10){\ft$2$}
\multiframe(0,-17)(10.5,0){1}(10,26){$ $} \put(4,-10){$\vdots$}
\multiframe(0,-27.5)(10.5,0){1}(10,10){\ft$n$}
\end{picture}
=~~~~ 
\begin{picture}(30,30)(0,0)
\multiframe(0,20)(10.5,0){1}(10,10){\ft$1$}
\multiframe(0,9.5)(10.5,0){1}(10,10){\ft$2$}
\multiframe(0,-17)(10.5,0){1}(10,26){$ $} \put(4,-10){$\vdots$}
\multiframe(0,-27.5)(10.5,0){1}(10,10){\ft$\partial$}
\end{picture}
.\vspace{1cm}
\ee 
Substituting this result in the decomposition (\ref{decom1})
yields 
\be
\label{momo} 
\begin{picture}(30,30)(0,0)
\multiframe(0,10)(10.5,0){1}(17,10){\ft$1$}
\multiframe(17.25,10)(10.5,0){1}(17,10){\ft$n\!\!+\!\!1$}
\multiframe(0,-0.5)(10.5,0){1}(17,10){\ft$2$}
\multiframe(0,-17)(10.5,0){1}(17,16){$ $} \put(7,-13.5){$\vdots$}
\multiframe(0,-27.5)(10.5,0){1}(17,10){\ft$n\!\!-\!\!1$}
\multiframe(0,-38)(10.5,0){1}(17,10){\ft$n$}
\end{picture}
\vspace{1.5cm}
\hspace*{2mm}\simeq~  
\begin{picture}(30,30)(0,0)
\multiframe(0,10)(10.5,0){1}(17,10){\ft$1$}
\multiframe(17.25,10)(10.5,0){1}(17,10){\ft$\partial$}
\multiframe(0,-0.5)(10.5,0){1}(17,10){\ft$2$}
\multiframe(0,-17)(10.5,0){1}(17,16){$ $} \put(7,-13.5){$\vdots$}
\multiframe(0,-27.5)(10.5,0){1}(17,10){\ft$n\!\!-\!\!1$}
\multiframe(0,-38)(10.5,0){1}(17,10){\ft$\partial$}
\end{picture}
\hspace*{2mm}\oplus\hspace*{2mm} 
\begin{picture}(30,30)(0,0)
\multiframe(0,10)(10.5,0){1}(17,10){\ft$1$}
\multiframe(17.25,10)(10.5,0){1}(17,10){\ft$n$}
\multiframe(0,-0.5)(10.5,0){1}(17,10){\ft$2$}
\multiframe(0,-17)(10.5,0){1}(17,16){$ $} \put(7,-13.5){$\vdots$}
\multiframe(0,-27.5)(10.5,0){1}(17,10){\ft$n\!\!-\!\!1$}
\multiframe(0,-38)(10.5,0){1}(17,10){\ft$\partial$}
\end{picture}
\hspace*{3mm}\longrightarrow\hspace*{2mm} 
\begin{picture}(30,30)(0,0)
\multiframe(0,10)(10.5,0){1}(17,10){\ft$1$}
\multiframe(17.25,10)(10.5,0){1}(17,10){\ft$n$}
\multiframe(0,-0.5)(10.5,0){1}(17,10){\ft$2$}
\multiframe(0,-17)(10.5,0){1}(17,16){$ $} \put(7,-13.5){$\vdots$}
\multiframe(0,-27.5)(10.5,0){1}(17,10){\ft$n\!\!-\!\!1$}
\multiframe(0,-38)(10.5,0){1}(17,10){\ft$\partial$}
\end{picture}
\hspace*{3mm}
\ee where the arrow indicates that we performed a field
redefinition. Thus, without loss of generality, the
right-hand-side can be assumed to contain a partial derivative in
the first column. With this preliminary result, the second
condition expressed in (\ref{cond1}), 
\be d^{\{2\}}
\hspace*{2mm}\vspace{.8cm} 
\begin{picture}(30,30)(0,0)
\multiframe(0,10)(10.5,0){1}(10,10){\ft$1$}
\multiframe(10.5,10)(10.5,0){1}(17,10){\ft$n\!\!+\!\!1$}
\multiframe(0,-0.5)(10.5,0){1}(10,10){\ft$2$}
\multiframe(0,-18)(10.5,0){1}(10,17){$ $} \put(4,-13.5){$\vdots$}
\multiframe(0,-28.5)(10.5,0){1}(10,10){\ft$n$}
\end{picture}
\equiv\hspace*{2mm}
\begin{picture}(25,30)(0,0)
\multiframe(0,10)(10.5,0){1}(10,10){\ft$1$}
\multiframe(10.5,10)(10.5,0){1}(17,10){\ft$n\!\!+\!\!1$}
\multiframe(10.5,-0.5)(10.5,0){1}(17,10){\ft$\partial$}
\multiframe(0,-0.5)(10.5,0){1}(10,10){\ft$2$}
\multiframe(0,-18)(10.5,0){1}(10,17){$ $} \put(4,-13.5){$\vdots$}
\multiframe(0,-28.5)(10.5,0){1}(10,10){\ft$n$}\put(54,-1){,}
\end{picture}
~~=0
\ee gives
\be\vspace{1.5cm}  
\begin{picture}(30,30)(0,0)
\multiframe(0,20)(10.5,0){1}(17,10){\ft$1$}
\multiframe(17.5,20)(10.5,0){1}(17,10){\ft$n$}
\multiframe(0,9.5)(10.5,0){1}(17,10){\ft$2$}
\multiframe(17.5,9.5)(10.5,0){1}(17,10){\ft$\partial$}
\put(7,-4){$\vdots$}
\multiframe(0,-10)(10.5,0){1}(17,19){$ $}
\multiframe(0,-20.5)(10.5,0){1}(17,10){\ft$n\!\!-\!\!1$}
\multiframe(0,-31)(10.5,0){1}(17,10){\ft$\partial$}
\end{picture}
\hspace*{4mm}=~0\,.
\label{mouche} \ee
The Poincar\'e lemma on the second column leads
to 
\be\vspace{1.5cm}  
\begin{picture}(30,30)(0,0)
\multiframe(0,10)(10.5,0){1}(17,10){\ft$1$}
\multiframe(17.5,10)(10.5,0){1}(17,10){\ft$n$}
\multiframe(0,-0.5)(10.5,0){1}(17,10){\ft$2$} \put(7,-14){$\vdots$}
\multiframe(0,-20)(10.5,0){1}(17,19){$ $}
\multiframe(0,-30.5)(10.5,0){1}(17,10){\ft$n\!\!-\!\!1$}
\multiframe(0,-41)(10.5,0){1}(17,10){\ft$\partial$}
\end{picture}
\hspace*{4mm}\simeq~~ 
\begin{picture}(20,30)(0,0)
\multiframe(0,-0.5)(10.5,0){1}(17,10){\ft$\partial$}
\end{picture}
\otimes\hspace*{1mm}  
\begin{picture}(30,30)(0,0)
\multiframe(0,10)(10.5,0){1}(17,10){\ft$1$}
\multiframe(0,-0.5)(10.5,0){1}(17,10){\ft$2$} \put(7,-14){$\vdots$}
\multiframe(0,-20)(10.5,0){1}(17,19){$ $}
\multiframe(0,-30.5)(10.5,0){1}(17,10){\ft$n\!\!-\!\!1$}
\multiframe(0,-40.5)(10.5,0){1}(17,10){\ft$n$}
\end{picture}
\simeq~~  
\begin{picture}(25,25)(0,0)
\multiframe(0,10)(10.5,0){1}(17,10){\ft$1$}
\multiframe(0,-0.5)(10.5,0){1}(17,10){\ft$2$} \put(7,-14){$\vdots$}
\multiframe(0,-20)(10.5,0){1}(17,19){$ $}
\multiframe(0,-30.5)(10.5,0){1}(17,10){\ft$n\!\!-\!\!1$}
\multiframe(0,-40.5)(10.5,0){1}(17,10){\ft$n$}
\multiframe(0,-51)(10.5,0){1}(17,10){\ft$\partial$}
\end{picture}
\oplus\hspace*{2mm}  
\begin{picture}(30,30)(0,0)
\multiframe(0,10)(10.5,0){1}(17,10){\ft$1$}
\multiframe(17.5,10)(10.5,0){1}(17,10){\ft$\partial$}
\multiframe(0,-0.5)(10.5,0){1}(17,10){\ft$2$} \put(7,-14){$\vdots$}
\multiframe(0,-20)(10.5,0){1}(17,19){$ $}
\multiframe(0,-30.5)(10.5,0){1}(17,10){\ft$n\!\!-\!\!1$}
\multiframe(0,-41)(10.5,0){1}(17,10){\ft$n$}
\end{picture}
~~~.
\label{moucheron} \ee 
The first totally antisymmetric component
vanishes since there is no component with the same symmetry on the
left-hand-side, implying that 
\be\vspace{1.4cm} 
\begin{picture}(30,30)(0,0)
\multiframe(0,10)(10.5,0){1}(17,10){\ft$1$}
\multiframe(0,-0.5)(10.5,0){1}(17,10){\ft$2$} \put(7,-14){$\vdots$}
\multiframe(0,-20)(10.5,0){1}(17,19){$ $}
\multiframe(0,-30.5)(10.5,0){1}(17,10){\ft$n\!\!-\!\!1$}
\multiframe(0,-41)(10.5,0){1}(17,10){\ft$n$}
\end{picture}
= ~~\;\; 
\begin{picture}(30,30)(0,0)
\multiframe(0,10)(10.5,0){1}(17,10){\ft$1$}
\multiframe(0,-0.5)(10.5,0){1}(17,10){\ft$2$} \put(7,-14){$\vdots$}
\multiframe(0,-20)(10.5,0){1}(17,19){$ $}
\multiframe(0,-30.5)(10.5,0){1}(17,10){\ft$n\!\!-\!\!1$}
\multiframe(0,-41)(10.5,0){1}(17,10){\ft$\partial$}
\end{picture}
\ee 
which in turn, substituted into (\ref{moucheron}), gives
\be\label{momotu}\vspace{1.5cm} 
\begin{picture}(30,30)(0,0)
\multiframe(0,10)(10.5,0){1}(17,10){\ft$1$}
\multiframe(17.5,10)(10.5,0){1}(17,10){\ft$n$}
\multiframe(0,-0.5)(10.5,0){1}(17,10){\ft$2$} \put(7,-14){$\vdots$}
\multiframe(0,-20)(10.5,0){1}(17,19){$ $}
\multiframe(0,-30.5)(10.5,0){1}(17,10){\ft$n\!\!-\!\!1$}
\multiframe(0,-41)(10.5,0){1}(17,10){\ft$\partial$}
\end{picture}
\hspace*{2mm}=\hspace*{2mm} 
\begin{picture}(30,30)(0,0)
\multiframe(0,10)(10.5,0){1}(17,10){\ft$1$}
\multiframe(17.5,10)(10.5,0){1}(17,10){\ft$\partial$}
\multiframe(0,-0.5)(10.5,0){1}(17,10){\ft$2$} \put(7,-14){$\vdots$}
\multiframe(0,-20)(10.5,0){1}(17,19){$ $}
\multiframe(0,-30.5)(10.5,0){1}(17,10){\ft$n\!\!-\!\!1$}
\multiframe(0,-41)(10.5,0){1}(17,10){\ft$\partial$}
\end{picture}
\quad .
\ee 
Substituting this result in (\ref{momo}) proves (\ref{resu1}).
%
\subsubsection{Second cohomology group}
%
We now turn to the proof that (\ref{cond2}) implies (\ref{resu2}).
The condition  (\ref{cond2}) reads 
\be\vspace{1.5cm} 
\begin{picture}(30,30)(0,0)
\multiframe(0,10)(10.5,0){1}(17,10){\ft$1$}
\multiframe(17.5,10)(10.5,0){1}(17,10){\ft$n\!\!+\!\!1$}
\multiframe(0,-0.5)(10.5,0){1}(17,10){\ft$2$}
\multiframe(17.5,-0.5)(10.5,0){1}(17,10){\ft$\partial$}
\put(7,-15){$\vdots$} \multiframe(0,-20)(10.5,0){1}(17,19){$ $}
\multiframe(0,-30.5)(10.5,0){1}(17,10){\ft$n$}
\multiframe(0,-41)(10.5,0){1}(17,10){\ft$\partial$}
\end{picture}
\hspace*{4mm}=~0\,
\ee 
whose type was already encountered in (\ref{mouche}) above. We
use our previous result (\ref{momotu}) and write 
\be\vspace{1.5cm}
\begin{picture}(30,30)(0,0)
\multiframe(0,10)(10.5,0){1}(17,10){\ft$1$}
\multiframe(17.5,10)(10.5,0){1}(17,10){\ft$n\!\!+\!\!1$}
\multiframe(0,-0.5)(10.5,0){1}(17,10){\ft$2$} \put(7,-15){$\vdots$}
\multiframe(0,-20)(10.5,0){1}(17,19){$ $}
\multiframe(0,-30.5)(10.5,0){1}(17,10){\ft$n$}
\multiframe(0,-41)(10.5,0){1}(17,10){\ft$\partial$}
\end{picture}
\hspace*{2mm}=\hspace*{2mm} 
\begin{picture}(30,30)(0,0)
\multiframe(0,10)(10.5,0){1}(17,10){\ft$1$}
\multiframe(17.5,10)(10.5,0){1}(17,10){\ft$\partial$}
\multiframe(0,-0.5)(10.5,0){1}(17,10){\ft$2$} \put(7,-15){$\vdots$}
\multiframe(0,-20)(10.5,0){1}(17,19){$ $}
\multiframe(0,-30.5)(10.5,0){1}(17,10){\ft$n$}
\multiframe(0,-41)(10.5,0){1}(17,10){\ft$\partial$}
\end{picture}
\ee 
or 
\be\vspace{1.5cm}  
\begin{picture}(30,30)(0,0)
\multiframe(0,10)(10.5,0){1}(17,10){\ft$1$}
\multiframe(17.5,10)(10.5,0){1}(17,10){\ft$n\!\!+\!\!1$}
\multiframe(0,-0.5)(10.5,0){1}(17,10){\ft$2$} \put(7,-15){$\vdots$}
\multiframe(0,-20)(10.5,0){1}(17,19){$ $}
\multiframe(0,-30.5)(10.5,0){1}(17,10){\ft$n$}
\multiframe(0,-41)(10.5,0){1}(17,10){\ft$\partial$}
\end{picture}
\hspace*{2mm}-\hspace*{2mm} 
\begin{picture}(30,30)(0,0)
\multiframe(0,10)(10.5,0){1}(17,10){\ft$1$}
\multiframe(17.5,10)(10.5,0){1}(17,10){\ft$\partial$}
\multiframe(0,-0.5)(10.5,0){1}(17,10){\ft$2$} \put(7,-15){$\vdots$}
\multiframe(0,-20)(10.5,0){1}(17,19){$ $}
\multiframe(0,-30.5)(10.5,0){1}(17,10){\ft$n$}
\multiframe(0,-41)(10.5,0){1}(17,10){\ft$\partial$}
\end{picture}
~=~0~.
\ee 
This kind of equation was also found before, in (\ref{cond1}),
when i=1. Then we are able to write 
\be\vspace{1.5cm}
\begin{picture}(30,30)(0,0)
\multiframe(0,10)(10.5,0){1}(17,10){\ft$1$}
\multiframe(17.5,10)(10.5,0){1}(17,10){\ft$n\!\!+\!\!1$}
\multiframe(0,-0.5)(10.5,0){1}(17,10){\ft$2$} \put(7,-15){$\vdots$}
\multiframe(0,-20)(10.5,0){1}(17,19){$ $}
\multiframe(0,-30.5)(10.5,0){1}(17,10){\ft$n$}
\end{picture}
\hspace*{2mm}-\hspace*{2mm} 
\begin{picture}(30,30)(0,0)
\multiframe(0,10)(10.5,0){1}(17,10){\ft$1$}
\multiframe(17.5,10)(10.5,0){1}(17,10){\ft$\partial$}
\multiframe(0,-0.5)(10.5,0){1}(17,10){\ft$2$} \put(7,-15){$\vdots$}
\multiframe(0,-20)(10.5,0){1}(17,19){$ $}
\multiframe(0,-30.5)(10.5,0){1}(17,10){\ft$n$}
\end{picture}
~=~ 
\begin{picture}(30,30)(0,0)
\multiframe(0,10)(10.5,0){1}(17,10){\ft$1$}
\multiframe(17.5,10)(10.5,0){1}(17,10){\ft$n$}
\multiframe(0,-0.5)(10.5,0){1}(17,10){\ft$2$} \put(7,-15){$\vdots$}
\multiframe(0,-20)(10.5,0){1}(17,19){$ $}
\multiframe(0,-30.5)(10.5,0){1}(17,10){\ft$\partial$}
\end{picture}
\ee 
which is the analogue of (\ref{momo}). Equivalently,
\be\vspace{1.5cm} 
\begin{picture}(30,30)(0,0)
\multiframe(0,10)(10.5,0){1}(17,10){\ft$1$}
\multiframe(17.5,10)(10.5,0){1}(17,10){\ft$n\!\!+\!\!1$}
\multiframe(0,-0.5)(10.5,0){1}(17,10){\ft$2$} \put(7,-15){$\vdots$}
\multiframe(0,-20)(10.5,0){1}(17,19){$ $}
\multiframe(0,-30.5)(10.5,0){1}(17,10){\ft$n$}
\end{picture}
\hspace*{2mm}=\hspace*{2mm} 
\begin{picture}(30,30)(0,0)
\multiframe(0,10)(10.5,0){1}(17,10){\ft$1$}
\multiframe(17.5,10)(10.5,0){1}(17,10){\ft$\partial$}
\multiframe(0,-0.5)(10.5,0){1}(17,10){\ft$2$} \put(7,-15){$\vdots$}
\multiframe(0,-20)(10.5,0){1}(17,19){$ $}
\multiframe(0,-30.5)(10.5,0){1}(17,10){\ft$n$}
\end{picture}
~+~ 
\begin{picture}(30,30)(0,0)
\multiframe(0,10)(10.5,0){1}(17,10){\ft$1$}
\multiframe(17.5,10)(10.5,0){1}(17,10){\ft$n\!\!+\!\!1$}
\multiframe(0,-0.5)(10.5,0){1}(17,10){\ft$2$} \put(7,-15){$\vdots$}
\multiframe(0,-20)(10.5,0){1}(17,19){$ $}
\multiframe(0,-30.5)(10.5,0){1}(17,10){\ft$\partial$}
\end{picture}
\ee 
which is the desired result.
%
\subsection{Generalized cohomology in $\Omega^{(*,*)}_{(2)}({\mathbb R}^D)$}
\label{subA2}
%

Here we proceed by induction on the number of boxes in the last
(second) column.
We will temporarily leave the diagrammatic exposition. For an
easier understanding of the following propositions, we sketch a
pictorial translation of the proof that
$H^{(n,l)}_{(1)}(\Omega_2({\mathbb R}^D)) \simeq 0\,,~~0<l<n<D$ in
subsection \ref{Diagrammatically}.
\\
{\it 
\noindent {\bf Induction hypothesis ${\cal S}_\ell$ :} Suppose that the three
following statements hold :
\ba
\bullet &&d^{\{1\}}\m(l_1,l_2)=0\,~~
\Rightarrow ~~\m(l_1,l_2) = d^{\{1\}}\n(l_1-1,l_2)\,,
\label{indu1}
\\
\bullet &&H^{(l_1,l_2)}_{(1)}(\Omega_{(2)}({\mathbb R}^D))\cong 0\,,
\label{indu2}
\\
\bullet &&H^{(l_1,l_2)}_{(2)}(\Omega_{(2)}({\mathbb R}^D))\cong 0\,,
\label{indu3}
\ea
where $0< l_1<D$, $0<l_2< \ell\leq  l_1$ and where
the notation $\m(l_1,l_2)$ indicates that
$\m\in\Omega^{(l_1,l_2)}_{(2)}({\mathbb R}^D)$, similarly
$\n\in\Omega^{(l_1-1,l_2)}_{(2)}({\mathbb R}^D)$.
The integer $\ell$ is fixed and is our induction parameter.}

The induction hypothesis ${\cal S}_{\ell}$ is that one knows the
cohomology of $d^{\{1\}}$ and the generalized cohomology for all
tensors whose second column has length strictly smaller than $\ell$.
What we showed in the above section \ref{subA1}
constitutes the ``initial conditions ${\cal S}_2$" of our
induction proof. The Poincar\'e lemma actually constitutes  ${\cal S}_1$.

To prove ${\cal S}_{\ell} \Rightarrow {\cal S}_{\ell+1}$ amounts to show
that we have the three assertions
(\ref{indu1}), (\ref{indu2}) and (\ref{indu3})
with the {\it new} conditions $0< l_1<D$, $0<l_2\leq  \ell \leq l_1$, {\it i.e.}
where the second column is now allowed to have length $l_2=\ell$.
\footnote{The case where the fixed induction parameter satisfies $\ell =l_1$
is a little bit particular, so will have to be treated separately.}

These three assertions (with the new conditions on the lengths of the columns) 
are proved in the following and lead to the result that
$H^{(l_1,l_2)}_{(*)}(\Omega_{(2)}({\mathbb R}^D))\simeq 0$
{\it{for any}} $(l_1,l_2)\in {\mathbb Y}^{(2)}$.

Before starting these three proofs and for later purposes, we introduce a total
order relation in the space $\Omega_{(S)}({\mathbb R}^D)$, naturally induced by
the total order relation (\ref{Ygrading}) for ${\mathbb Y}^{(S)}$.
\\
If $\a(l_1,\ldots,l_S)$ and $\b(l'_1,\ldots,l'_{S})$ belong to
$\Omega_{(S)}^{(l_1,\ldots,l_S)}({\mathbb R}^D)$ and
$\Omega_{(S)}^{(l'_1,\ldots,l'_{S})}({\mathbb R}^D)$, respectively, then
\be
\a(l_1,\ldots,l_S) \ll \b(l'_1,\ldots,l'_{S})
\ee if and only if
\ba
l_k&=&l'_k\,, ~~1\leq k\leq K\,, ~~{\rm{and}}\nonumber \\
l_{K+1}&\leq& l'_{K+1}\,, \ea where $K$ is an integer satisfying
$1\leq K \leq S$. This lexicographic ordering induces a grading in 
$\Omega_{(S)}({\mathbb R}^D)$, that we call
``L-grading''.
This L-grading is a generalization to $\Omega_{(S)}({\mathbb R}^D)$ of the
form-grading
in $\Omega_{(1)}({\mathbb R}^D)$.
In the sequel we will use hatted symbols to denote
multiforms of $\Omega_{[S]}({\mathbb R}^D)$, while the unhatted
tensors belong to $\Omega_{(S)}({\mathbb R}^D)$.
\\

Turning back to our inductive proof, we begin with the
\begin{lemma}
\label{lemma1} 
If $\cs_{\ell}$ is satisfied, then
\be d^{\{1\}}\m(l_1,\ell)=0\,, ~0< l_1<D\,\,
\Longrightarrow \,\,
\m(l_1,\ell)=d^{\{1\}}\n(l_1-1,\ell). \label{2col}\ee
\end{lemma}
\proof{(1) $\ell< l_1$. \\ \noindent Applying the Poincar\'e
lemma to the cocycle condition in Eqn (\ref{2col}), viewing the second 
column as a spectator, yields 
$\m(l_1,\ell)\simeq d_1 \hat{\n}(l_1-1,\ell)$, where
$\hat{\n}(l_1-1,\ell)\in\O_{[2]}^{l_1-1,\ell}$. Decomposing the
right-hand-side (expressed in terms of multiforms) into irrep. of
$GL(D,\mathbb R)$ gives 
\ba 
\m(l_1,\ell)&\simeq&
d^{\{1\}}\n(l_1-1,\ell)+d^{\{2\}}\n(l_1,\ell-1)+d^{\{1\}}\n(l_1,\ell-1)\nn\\
&&+d^{\{2\}}\n(l_1+1,\ell-2)+(\ldots)\,, 
\ea 
where $(\ldots)$
denotes tensors of higher L-grading. Because the third and fourth
terms do not belong to $\O_{(2)}^{(l_1,\ell)}$, they must cancel :
\be 
d^{\{1\}}\n(l_1,\ell-1)+d^{\{2\}}\n(l_1+1,\ell-2)=0\,.
\label{this} 
\ee 
Applying the operator $d^{\{2\}}$ to (\ref{this})
gives $d^{\{1,2\}}\n(l_1,\ell-1)=0$. The induction hypothesis
${\cal S}_{\ell}$ allows us to write
$\n(l_1,\ell-1)=d^{\{1\}}\n(l_1-1,\ell-1)+d^{\{2\}}\n(l_1,\ell-2)$.
Substituting back in the decomposition of $\m(l_1,\ell)$, we find,
after the redefinition 
$\n'(l_1-1,\ell)=\n(l_1-1,\ell)+d^{\{2\}}\n(l_1-1,\ell-1)$, the result we were
looking for : 
\be \m(l_1,\ell)=d^{\{1\}}\n'(l_1-1,\ell)\,.
\label{resulem2col} 
\ee

(2) The case $\ell=l_1$ can be analyzed in the same way. \\
\noindent The equation 
\be 
d^{\{1\}}\m(l_1,l_1)=0\,,~~0<l_1<D 
\ee
gives, after applying the standard Poincar\'e lemma, that
$\m(l_1,l_1)\simeq d_1 \hat{\n}(l_1-1,l_1)$,
$\hat{\n}(l_1-1,l_1)\in\O_{[2]}^{l_1-1,l_1}\simeq\O_{[2]}^{l_1,l_1-1}$.
In terms of irrep. of $GL(D,\mathbb R)$, we get $\m(l_1,l_1)\simeq
d^{\{2\}}
\n(l_1,l_1-1)+d^{\{1\}}\n(l_1,l_1-1)+d^{\{2\}}\n(l_1+1,l_1-2)+(\ldots)$
where $(\ldots)$ denote terms of higher L-grading. The sum of the
second and third terms must vanish, and applying $d^{\{2\}}$ gives
$d^{\{1,2\}}\n(l_1,l_1-1)=0$. By virtue of our hypothesis of induction 
${\cal S}_{\ell}$, we obtain
$\n(l_1,l_1-1)=d^{\{1\}}\n(l_1-1,l_1-1)+d^{\{2\}}\n(l_1,l_1-2)$.
Here, the result which emerges after substituting the above
equation in the decomposition of $\m(l_1,l_1)$ and preforming a field
redefinition, is \be \m(l_1,l_1)=d^{\{1,2\}}\n(l_1-1,l_1-1)\,.
\label{resulem2idemcol} \ee } Note that the case $\ell = l_1$ gave
us for free
\begin{proposition}
\label{theoHll1} If $\cs_{\ell}$ is satisfied, then
$H^{(\ell,\ell)}_{(1)}(\Omega_{(2)}({\mathbb
R}^D))\cong 0$\,.
\end{proposition}
\noindent Having the Lemma \ref{lemma1} at our disposal, we now
proceed to prove
\begin{proposition}
\label{Hnl} If $\cs_{\ell}$ is satisfied, then
$H^{(l_1,\ell)}_{(1)}(\Omega_{(2)}({\mathbb
R}^D))\cong 0\,$.
\end{proposition}
\noindent It states that the cocycle conditions 
\be
d^{\{i\}}\m(l_1,\ell)=0\,,~~i\in\{1,2\},~~0< l_1<D
\label{cocycle1} 
\ee imply that 
\be
\m(l_1,\ell)=d^{\{1,2\}}\n(l_1-1,\ell-1)\,. 
\ee 
\proof{(1)
$\ell< l_1$. \\ \noindent In the case $i=1$, the conditions
(\ref{cocycle1}) give, using Lemma \ref{lemma1}, that \be
\m(l_1,\ell)=d^{\{1\}}\n(l_1-1,\ell). \label{intermed1} \ee
Substituting this into the condition (\ref{cocycle1}) for $i=2$
yields \be d^{\{1,2\}}\n(l_1-1,\ell)=0. \label{cocycle12} \ee
Using the Poincar\'e lemma on the second column, we have \ba
d^{\{1\}}\n(l_1-1,\ell)&\simeq& d_1
\hat{\n}(l_1-1,\ell)\nn\\&\simeq&
d^{\{2\}}\n(l_1,\ell-1)+d^{\{1\}}\n(l_1,\ell-1)\nn\\&&+d^{\{2\}}
\n(l_1+1,\ell-2) +(\ldots)\,,\ea where, as before, we used the
branching rules for $GL(D,\mathbb R)$ and $(\ldots)$ corresponds
to terms of higher L-grading. The sum of the second and third
terms of the right-hand side must vanish, as does the action of
$d^{\{2\}}$ on it. As a consequence,
$\n(l_1,\ell-1)=d^{\{1\}}\n(l_1-1,\ell-1)+d^{\{2\}}\n(l_1,\ell-2)$,
hence $d^{\{1\}}\n(l_1-1,\ell)=d^{\{1,2\}}\n(l_1-1,\ell-1)$.
Substituting this into (\ref{intermed1}) we finally have \be
\m(l_1,\ell)=d^{\{1,2\}}\n(l_1-1,\ell-1). \ee which proves the
proposition.

(2) $\ell =l_1$. \\ \noindent
This case was already obtained in Proposition \ref{theoHll1}.
}

The following vanishing of cohomology still remains to be shown:
\begin{proposition}
\label{Hnlbis}
If $\cs_{\ell}$ is satisfied, then 
$H^{(l_1,\ell)}_{(2)}(\Omega_{(2)}({\mathbb R}^D))\cong 0\,.$
\end{proposition}
\proof{The cocycle condition with $\ell < l_1$ has in fact
already been encountered in (\ref{cocycle12}).
We can then use the results already obtained in the proof of the
Proposition \ref{Hnl} to write that the cocycle condition
\be
d^{\{1,2\}}\m(l_1,\ell)=0\,, \quad 0<l_1<D, ~\ell \neq l_1
\ee
leads to $d^{\{1\}}\m(l_1,\ell)=d^{\{1,2\}}\m(l_1,\ell-1)$.
Rewriting this equation as
$d^{\{1\}}[\m(l_1,\ell)-d^{\{2\}}\m(l_1,\ell-1)]=0$ and using the
results of Lemma \ref{lemma1} we obtain
$\m(l_1,\ell)-d^{\{2\}}\m(l_1,\ell-1)= d^{\{1\}}\m(l_1-1,\ell)$,
{\it{i.e.}}
\be
\m(l_1,\ell)=d^{\{2\}}\m(l_1,\ell-1)+d^{\{1\}}\m(l_1-1,\ell).
\label{cocycle1+2}
\ee
Had we started with the cocycle condition
$d^{\{1,2\}}\m(l_1,\ell=l_1)=0$, $0<l_1<D$, we would have found
$d^{\{1\}}\m(l_1,l_1)=d^{\{1,2\}}\m(l_1,l_1-1)$, then
$\m(l_1,l_1)-d^{\{2\}}\m(l_1,l_1-1)=d^{\{1,2\}}\m(l_1-1,l_1-1)$,
and after a field redefinition, the result
\be
\m(l_1,l_1)=d^{\{2\}}\m(l_1,l_1-1)
\ee
which is the coboundary
condition analogous to (\ref{cocycle1+2}) in the case of maximally
filled tensors in $\O_{2}({\mathbb R}^D)$.}

\noindent{\bf{Conclusions}}
\\
Our inductive proof provided us with the following results about
the generalized cohomologies of $d^{\{i\}}$, $i\in\{1,2\}$, in the
space $\O_{(2)}$ : 
\ba H^{(l_1,l_2)}_{(*)}(\Omega_{(2)}({\mathbb R}^D))\cong 0\,,
~\forall (l_1,l_2)\in {\mathbb Y}^{(2)}, l_2 \neq 0. 
\ea
%
\subsection{Diagrammatical presentation}\label{Diagrammatically}
\label{subA3}
%
The pictorial translation of {Lemma \ref{lemma1}}, 
which proved to be crucial in proving 
{$H^{(n,l)}_{(1)}(\Omega_{(2)}({\mathbb R}^D))\cong 0\,$}, reads
\be\vspace{3cm}  
\begin{picture}(30,30)(0,0)
\multiframe(0,10)(10.5,0){1}(10,10){\ft$ 1$} \put(4,-10){$\vdots$}
\multiframe(0,-20)(10.5,0){1}(10,29.5){$ $}
\multiframe(10.5,10)(10.5,0){1}(10,10){\ft$1 $}
\multiframe(10.5,-9.5)(10.5,0){1}(10,19){$ $}
\multiframe(10.5,-20)(10.5,0){1}(10,10){\ft$l$}
\multiframe(0,-39.5)(10.5,0){1}(10,19){$ $}
\put(4,-34.5){$\vdots$} \put(14,-5){$\vdots$}
\multiframe(0,-49.5)(10.5,0){1}(10,10){\ft$n$}
\multiframe(0,-60)(10.5,0){1}(10,10){\ft$\partial$}
\end{picture}
=0~~~~\Rightarrow~~ 
\begin{picture}(30,30)(0,0)
\multiframe(0,10)(10.5,0){1}(10,10){\ft$ 1$} \put(4,-10){$\vdots$}
\multiframe(0,-20)(10.5,0){1}(10,29.5){$ $}
\multiframe(10.5,10)(10.5,0){1}(10,10){\ft$1 $}
\multiframe(10.5,-9.5)(10.5,0){1}(10,19){$ $}
\multiframe(10.5,-20)(10.5,0){1}(10,10){\ft$l$}
\multiframe(0,-39.5)(10.5,0){1}(10,19){$ $}
\put(4,-34.5){$\vdots$} \put(14,-5){$\vdots$}
\multiframe(0,-49.5)(10.5,0){1}(10,10){\ft$n$}
\end{picture}
~=~~~
\begin{picture}(30,30)(0,0)
\multiframe(0,10)(10.5,0){1}(10,10){\ft$ 1$} \put(4,-10){$\vdots$}
\multiframe(0,-20)(10.5,0){1}(10,29.5){$ $}
\multiframe(10.5,10)(10.5,0){1}(10,10){\ft$1 $}
\multiframe(10.5,-9.5)(10.5,0){1}(10,19){$ $}
\multiframe(10.5,-20)(10.5,0){1}(10,10){\ft$l$}
\multiframe(0,-39.5)(10.5,0){1}(10,19){$ $}
\put(4,-34.5){$\vdots$} \put(14,-5){$\vdots$}
\multiframe(0,-49.5)(10.5,0){1}(10,10){\ft$\partial$}
\end{picture}
~.
\ee 
For practical purposes and for simplicity, we fix $l=2$. 
Using the standard Poincar\'e lemma, one obtains
\be\vspace{1cm} 
\begin{picture}(30,30)(0,0)
\multiframe(0,20)(10.5,0){1}(10,10){\ft$1$}
\multiframe(0,9.5)(10.5,0){1}(10,10){\ft$2$}
\multiframe(0,-8)(10.5,0){1}(10,17){$ $} \put(4,-5){$\vdots$}
\multiframe(0,-18.5)(10.5,0){1}(10,10){\ft$n$}
\multiframe(0,-29)(10.5,0){1}(10,10){\ft$\partial$}
\multiframe(10.5,20)(10.5,0){1}(10,10){\ft$*$}
\multiframe(10.5,9.5)(10.5,0){1}(10,10){\ft$*$}
\end{picture}
~=0~\Rightarrow~  
\begin{picture}(30,30)(0,0)
\multiframe(0,10)(10.5,0){1}(10,10){\ft$1$}
\multiframe(0,-0.5)(10.5,0){1}(10,10){\ft$2$}
\multiframe(0,-18)(10.5,0){1}(10,17){$ $} \put(4,-15){$\vdots$}
\multiframe(0,-28.5)(10.5,0){1}(10,10){\ft$n$}
\multiframe(10.5,10)(10.5,0){1}(10,10){\ft$*$}
\multiframe(10.5,-0.5)(10.5,0){1}(10,10){\ft$*$}
\end{picture}
\simeq~ 
\begin{picture}(15,30)(0,0)
\multiframe(0,0)(10.5,0){1}(10,10){\ft$\partial$}
\end{picture}
\otimes~\Big(~  
\begin{picture}(25,30)(0,0)
\multiframe(0,10)(10.5,0){1}(17,10){\ft$1$}
\multiframe(0,-0.5)(10.5,0){1}(17,10){\ft$2$}
\multiframe(0,-18)(10.5,0){1}(17,17){$ $} \put(6,-15){$\vdots$}
\multiframe(0,-28.5)(10.5,0){1}(17,10){\ft$n\!\!-\!\!1$}
\end{picture}
\otimes~  
\begin{picture}(15,30)(0,0)
\multiframe(0,5)(10.5,0){1}(10,10){\ft$*$}
\multiframe(0,-5.5)(10.5,0){1}(10,10){\ft$*$}
\end{picture}
\Big) \ee i.e. \be\vspace{1.8cm} 
\begin{picture}(30,30)(0,0)
\multiframe(0,10)(10.5,0){1}(17,10){\ft$1$}
\multiframe(0,-0.5)(10.5,0){1}(17,10){\ft$2$}
\multiframe(0,-18)(10.5,0){1}(17,17){$ $} \put(7,-15){$\vdots$}
\multiframe(0,-28.5)(10.5,0){1}(17,10){\ft$n\!\!-\!\!1$}
\multiframe(0,-39)(10.5,0){1}(17,10){\ft$n$}
\multiframe(17.5,10)(10.5,0){1}(10,10){\ft$* $}
\multiframe(17.5,-0.5)(10.5,0){1}(10,10){\ft$* $}
\end{picture}
\simeq~  
\begin{picture}(30,30)(0,0)
\multiframe(0,10)(10.5,0){1}(17,10){\ft$1$}
\multiframe(0,-0.5)(10.5,0){1}(17,10){\ft$2$}
\multiframe(0,-18)(10.5,0){1}(17,17){$ $} \put(7,-15){$\vdots$}
\multiframe(0,-28.5)(10.5,0){1}(17,10){\ft$n\!\!-\!\!1$}
\multiframe(0,-39)(10.5,0){1}(17,10){\ft$\partial$}
\multiframe(17.5,10)(10.5,0){1}(10,10){\ft$* $}
\multiframe(17.5,-0.5)(10.5,0){1}(10,10){\ft$* $}
\end{picture}
\oplus  
\begin{picture}(30,30)(0,0)
\multiframe(0,10)(10.5,0){1}(17,10){\ft$1$}
\multiframe(0,-0.5)(10.5,0){1}(17,10){\ft$2$}
\multiframe(0,-18)(10.5,0){1}(17,17){$ $} \put(7,-15){$\vdots$}
\multiframe(0,-28.5)(10.5,0){1}(17,10){\ft$n\!\!-\!\!1$}
\multiframe(0,-39)(10.5,0){1}(17,10){\ft$* $}
\multiframe(17.5,10)(10.5,0){1}(10,10){\ft$ *$}
\multiframe(17.5,-0.5)(10.5,0){1}(10,10){\ft$\partial$}
\end{picture}
\oplus  
\begin{picture}(30,30)(0,0)
\multiframe(0,20)(10.5,0){1}(17,10){\ft$1$}
\multiframe(0,9.5)(10.5,0){1}(17,10){\ft$2$}
\multiframe(0,-8)(10.5,0){1}(17,17){$ $} \put(7,-5){$\vdots$}
\multiframe(0,-18.5)(10.5,0){1}(17,10){\ft$n\!\!-\!\!1$}
\multiframe(0,-29)(10.5,0){1}(17,10){\ft$*$}
\multiframe(0,-39.5)(10.5,0){1}(17,10){\ft$\partial$}
\multiframe(17.5,20)(10.5,0){1}(10,10){\ft$* $}
\end{picture}
\oplus 
\begin{picture}(30,30)(0,0)
\multiframe(0,20)(10.5,0){1}(17,10){\ft$1$}
\multiframe(0,9.5)(10.5,0){1}(17,10){\ft$2$}
\multiframe(0,-8)(10.5,0){1}(17,17){$ $} \put(7,-5){$\vdots$}
\multiframe(0,-18.5)(10.5,0){1}(17,10){\ft$n\!\!-\!\!1$}
\multiframe(0,-29)(10.5,0){1}(17,10){\ft$* $}
\multiframe(0,-39.5)(10.5,0){1}(17,10){\ft$* $}
\multiframe(17.5,20)(10.5,0){1}(10,10){\ft$ \partial$}
\end{picture}
\oplus  
\begin{picture}(30,30)(0,0)
\multiframe(0,20)(10.5,0){1}(17,10){\ft$1$}
\multiframe(0,9.5)(10.5,0){1}(17,10){\ft$2$}
\multiframe(0,-8)(10.5,0){1}(17,17){$ $} \put(7,-5){$\vdots$}
\multiframe(0,-18.5)(10.5,0){1}(17,10){\ft$n\!\!-\!\!1$}
\multiframe(0,-29)(10.5,0){1}(17,10){\ft$* $}
\multiframe(0,-39.5)(10.5,0){1}(17,10){\ft$* $}
\multiframe(0,-50)(10.5,0){1}(17,10){\ft$ \partial$}
\end{picture}
~.
\label{debal1} \ee 
The condition that the sum of the third and
fourth terms of the right-hand side vanishes, implies, due to the
induction hypothesis, that the tensor in the second term can be
written as 
\be\vspace{1.5cm} 
\begin{picture}(30,30)(0,0)
\multiframe(0,10)(10.5,0){1}(17,10){\ft$1$}
\multiframe(0,-0.5)(10.5,0){1}(17,10){\ft$2$}
\multiframe(0,-18)(10.5,0){1}(17,17){$ $} \put(7,-15){$\vdots$}
\multiframe(0,-28.5)(10.5,0){1}(17,10){\ft$n\!\!-\!\!1$}
\multiframe(0,-39)(10.5,0){1}(17,10){\ft$*$}
\multiframe(17.5,10)(10.5,0){1}(10,10){\ft$*$}
\end{picture}
~=~  
\begin{picture}(30,30)(0,0)
\multiframe(0,10)(10.5,0){1}(17,10){\ft$1$}
\multiframe(0,-0.5)(10.5,0){1}(17,10){\ft$2$}
\multiframe(0,-18)(10.5,0){1}(17,17){} \put(7,-15){$\vdots$}
\multiframe(0,-28.5)(10.5,0){1}(17,10){\ft$n\!\!-\!\!1$}
\multiframe(0,-39)(10.5,0){1}(17,10){\ft$\partial$}
\multiframe(17.5,10)(10.5,0){1}(10,10){\ft$*$}
\end{picture}
~+~  
\begin{picture}(30,30)(0,0)
\multiframe(0,10)(10.5,0){1}(17,10){\ft$1$}
\multiframe(0,-0.5)(10.5,0){1}(17,10){\ft$2$}
\multiframe(0,-18)(10.5,0){1}(17,17){$ $} \put(7,-15){$\vdots$}
\multiframe(0,-28.5)(10.5,0){1}(17,10){\ft$n\!\!-\!\!1$}
\multiframe(0,-39)(10.5,0){1}(17,10){\ft$*$}
\multiframe(17.5,10)(10.5,0){1}(10,10){\ft$\partial$}
\end{picture}
\ee 
which, substituted into (\ref{debal1}), gives 
\be
\vspace{1cm} 
\begin{picture}(30,30)(0,0)
\multiframe(0,10)(10.5,0){1}(17,10){\ft$1$}
\multiframe(0,-0.5)(10.5,0){1}(17,10){\ft$2$}
\multiframe(0,-18)(10.5,0){1}(17,17){$ $} \put(7,-15){$\vdots$}
\multiframe(0,-28.5)(10.5,0){1}(17,10){\ft$n\!\!-\!\!1$}
\multiframe(0,-39)(10.5,0){1}(17,10){\ft$n$}
\multiframe(17.5,10)(10.5,0){1}(10,10){\ft$ *$}
\multiframe(17.5,-0.5)(10.5,0){1}(10,10){\ft$* $}
\end{picture}
~=~  
\begin{picture}(30,30)(0,0)
\multiframe(0,10)(10.5,0){1}(17,10){\ft$1$}
\multiframe(0,-0.5)(10.5,0){1}(17,10){\ft$2$}
\multiframe(0,-18)(10.5,0){1}(17,17){$ $} \put(7,-15){$\vdots$}
\multiframe(0,-28.5)(10.5,0){1}(17,10){\ft$n\!\!-\!\!1$}
\multiframe(0,-39)(10.5,0){1}(17,10){\ft$\partial$}
\multiframe(17.5,10)(10.5,0){1}(10,10){\ft$* $}
\multiframe(17.5,-0.5)(10.5,0){1}(10,10){\ft$ *$}
\end{picture}
~.
\ee
\\
Once Lemma \ref{lemma1} is obtained, we turn to the pictorial description 
of the proof for {$H^{(n,l)}_{(1)}(\Omega_{(2)}({\mathbb R}^D))\cong 0\,$}.
The first cocycle condition 
\be 
d^{\{1\}} \a(n,l)=0 
\ee 
is represented by
\be
\vspace{1cm} 
\begin{picture}(30,30)(0,0) 
\multiframe(0,35)(10.5,0){1}(10,10){\ft$1$} \put(4,15){$\vdots$} 
\multiframe(0,5)(10.5,0){1}(10,29.5){$ $}
\multiframe(10.5,35)(10.5,0){1}(10,10){\ft$ 1$}
\multiframe(10.5,15.5)(10.5,0){1}(10,19){$ $}
\multiframe(10.5,5)(10.5,0){1}(10,10){\ft$l$}
\multiframe(0,-14.5)(10.5,0){1}(10,19){$ $} \put(4,-9.5){$\vdots$}
\put(14,20){$\vdots$} \multiframe(0,-25)(10.5,0){1}(10,10){\ft$n$}
\multiframe(0,-35.5)(10.5,0){1}(10,10){\ft$\partial$}
\end{picture}
~=~0\,. \ee 
Using  Lemma \ref{lemma1}, its solution is (as we showed pictorially in the 
case where $l=2$), 
\be\vspace{1cm}  
\begin{picture}(30,30)(0,0)
\multiframe(0,20)(10.5,0){1}(10,10){\ft$1$} \put(4,0){$\vdots$}
\multiframe(0,-10)(10.5,0){1}(10,29.5){$ $}
\multiframe(10.5,20)(10.5,0){1}(10,10){\ft$1$}
\multiframe(10.5,0.5)(10.5,0){1}(10,19){$ $}
\multiframe(10.5,-10)(10.5,0){1}(10,10){\ft$l$}
\multiframe(0,-29.5)(10.5,0){1}(10,19){$ $}
\put(4,-24.5){$\vdots$} \put(14,5){$\vdots$}
\multiframe(0,-40)(10.5,0){1}(10,10){\ft$n$}
\end{picture}
~=~~~  
\begin{picture}(30,30)(0,0)
\multiframe(0,20)(10.5,0){1}(10,10){\ft$1$} \put(4,0){$\vdots$}
\multiframe(0,-10)(10.5,0){1}(10,29.5){$ $}
\multiframe(10.5,20)(10.5,0){1}(10,10){\ft$1$}
\multiframe(10.5,0.5)(10.5,0){1}(10,19){$ $}
\multiframe(10.5,-10)(10.5,0){1}(10,10){\ft$l$}
\multiframe(0,-29.5)(10.5,0){1}(10,19){$ $}
\put(4,-24.5){$\vdots$} \put(14,5){$\vdots$}
\multiframe(0,-40)(10.5,0){1}(10,10){\ft$\partial$}
\end{picture}
\,.
\label{inte2} 
\ee 
Substituting in the second cocycle condition 
\be
d^{\{2\}}\a(n,l)=0 
\ee 
yields 
\be\vspace{1.5cm} 
\begin{picture}(30,30)(0,0)
\multiframe(0,25)(10.5,0){1}(17,10){\ft$1$} \put(7,5){$\vdots$}
\multiframe(0,-5)(10.5,0){1}(17,29.5){$ $}
\multiframe(17.5,25)(10.5,0){1}(17,10){\ft$1$}
\multiframe(17.5,5.5)(10.5,0){1}(17,19){$ $}
\multiframe(17.5,-5)(10.5,0){1}(17,10){\ft$l$}
\multiframe(17.5,-15.5)(10.5,0){1}(17,10){\ft$\partial$}
\multiframe(0,-24.5)(10.5,0){1}(17,19){$ $}
\put(7,-19.5){$\vdots$} \put(24,10){$\vdots$}
\multiframe(0,-35)(10.5,0){1}(17,10){\ft$n\!\!-\!\!1$}
\multiframe(0,-45.5)(10.5,0){1}(17,10){\ft$\partial$}
\end{picture}
~~~=~0\,. 
\label{a(n+d,l+d)} 
\ee 
Applying the Poincar\'e lemma on
the second column, viewing the first one as a spectator, gives
\ba  
\begin{picture}(30,30)(0,0)
\multiframe(0,20)(10.5,0){1}(17,10){\ft$1$} \put(7,0){$\vdots$}
\multiframe(0,-10)(10.5,0){1}(17,29.5){$ $}
\multiframe(17.5,20)(10.5,0){1}(17,10){\ft$1$}
\multiframe(17.5,0.5)(10.5,0){1}(17,19){$ $}
\multiframe(17.5,-10)(10.5,0){1}(17,10){\ft$l$}
\multiframe(0,-29.5)(10.5,0){1}(17,19){$ $}
\put(7,-24.5){$\vdots$} \put(24,5){$\vdots$}
\multiframe(0,-40)(10.5,0){1}(17,10){\ft$n\!\!-\!\!1$}
\multiframe(0,-50.5)(10.5,0){1}(17,10){\ft$\partial$}
\end{picture}
~~&\simeq&~  
\begin{picture}(30,30)(0,0)
\multiframe(0,0)(10.5,0){1}(10,10){\ft$\partial$}
\end{picture}
\!\!\!\!\!\!\!\!
\otimes\,\,\Big(~~
\begin{picture}(30,30)(0,0)
\multiframe(0,10)(10.5,0){1}(17,10){\ft$1$}
\multiframe(0,-15.5)(10.5,0){1}(17,25){$ $} \put(7,-7.5){$\vdots$}
\multiframe(0,-26)(10.5,0){1}(17,10){\ft$n$}
\end{picture}
 \!\!\!\!\! \otimes 
\begin{picture}(30,30)(0,0)
\multiframe(0,10)(10.5,0){1}(17,10){\ft$1$}
\multiframe(0,-10.5)(10.5,0){1}(17,20){$ $} \put(7,-5.5){$\vdots$}
\multiframe(0,-21)(10.5,0){1}(17,10){\ft$l\!\!-\!\!1$}
\end{picture}
\Big)
\nonumber \\ \nonumber \\ \nonumber \\
~~&\simeq&~  
\begin{picture}(30,30)(0,0)
\multiframe(0,20)(10.5,0){1}(17,10){\ft$1$} \put(7,0){$\vdots$}
\multiframe(0,-10)(10.5,0){1}(17,29.5){$ $}
\multiframe(17.5,20)(10.5,0){1}(17,10){\ft$1$}
\multiframe(17.5,0.5)(10.5,0){1}(17,19){$ $}
\multiframe(17.5,-10)(10.5,0){1}(17,10){\ft$l\!\!-\!\!1$}
\multiframe(17.5,-20.5)(10.5,0){1}(17,10){\ft$\partial$}
\multiframe(0,-29.5)(10.5,0){1}(17,19){$ $}
\put(7,-24.5){$\vdots$} \put(24,5){$\vdots$}
\multiframe(0,-40)(10.5,0){1}(17,10){\ft$n$}
\end{picture}
~~\oplus  
\begin{picture}(30,30)(0,0)
\multiframe(0,20)(10.5,0){1}(17,10){\ft$1$} \put(7,0){$\vdots$}
\multiframe(0,-10)(10.5,0){1}(17,29.5){$ $}
\multiframe(17.5,20)(10.5,0){1}(17,10){\ft$1$}
\multiframe(17.5,0.5)(10.5,0){1}(17,19){$ $}
\multiframe(17.5,-10)(10.5,0){1}(17,10){\ft$l\!\!-\!\!1$}
\multiframe(0,-29.5)(10.5,0){1}(17,19){$ $}
\put(7,-24.5){$\vdots$} \put(24,5){$\vdots$}
\multiframe(0,-40)(10.5,0){1}(17,10){\ft$n$}
\multiframe(0,-50.5)(10.5,0){1}(17,10){\ft$\partial$}
\end{picture}
~~\oplus\,  
\begin{picture}(30,30)(0,0)
\multiframe(0,20)(10.5,0){1}(17,10){\ft$1$} \put(7,0){$\vdots$}
\multiframe(0,-10)(10.5,0){1}(17,29.5){$ $}
\multiframe(17.5,20)(10.5,0){1}(17,10){\ft$1$}
\multiframe(17.5,0.5)(10.5,0){1}(17,19){$ $}
\multiframe(17.5,0.5)(10.5,0){1}(17,10){\ft$l\!\!-\!\!2$}
\multiframe(17.5,-10)(10.5,0){1}(17,10){\ft$\partial$}
\put(24,16){.} \put(24,13){.}
\multiframe(0,-29.5)(10.5,0){1}(17,19){$ $}
\put(7,-24.5){$\vdots$} 
\multiframe(0,-40)(10.5,0){1}(17,10){\ft$n$}
\multiframe(0,-50.5)(10.5,0){1}(17,10){\ft$n\!\!+\!\!1$}
\end{picture}
~~\oplus\ldots \label{exp} 
\ea
\\
\\
\\
where the dots in the above equation correspond to tensors of
higher order $L$-grading (whose first column has a length greater
or equal to $n+2$). The second and third terms must cancel because
they do not have the symmetry of the left-hand side. Applying
$d^{\{2\}}$ on the sum of the second and third term and using our
hypothesis of induction, we obtain 
\be\vspace{1.5cm} 
\begin{picture}(30,30)(0,0)
\multiframe(0,20)(10.5,0){1}(17,10){\ft$1$} \put(7,-9){$\vdots$}
\multiframe(0,-25)(10.5,0){1}(17,44.5){$ $}
\multiframe(17.5,20)(10.5,0){1}(17,10){\ft$1$}
\multiframe(17.5,3.5)(10.5,0){1}(17,16){$ $}
\multiframe(17.5,-7)(10.5,0){1}(17,10){\ft$l\!\!-\!\!1$}
\multiframe(17.5,-17.5)(10.5,0){1}(17,10){\ft$l$}
\put(24,7){$\vdots$}
\multiframe(0,-35.5)(10.5,0){1}(17,10){\ft$n\!\!-\!\!1$}
\multiframe(0,-46)(10.5,0){1}(17,10){\ft$\partial$}
\end{picture}
~~\simeq~ 
\begin{picture}(30,30)(0,0)
\multiframe(0,20)(10.5,0){1}(17,10){\ft$1$} \put(7,-9){$\vdots$}
\multiframe(0,-25)(10.5,0){1}(17,44.5){$ $}
\multiframe(17.5,20)(10.5,0){1}(17,10){\ft$1$}
\multiframe(17.5,3.5)(10.5,0){1}(17,16){$ $}
\multiframe(17.5,-7)(10.5,0){1}(17,10){\ft$l\!\!-\!\!1$}
\multiframe(17.5,-17.5)(10.5,0){1}(17,10){\ft$\partial$}
\put(24,7){$\vdots$}
\multiframe(0,-35.5)(10.5,0){1}(17,10){\ft$n\!\!-\!\!1$}
\multiframe(0,-46)(10.5,0){1}(17,10){\ft$\partial$}
\end{picture}
~~~~.
\ee 
This, substituted back into (\ref{inte2}), gives us the
vanishing of $H^{(n,l)}_{(1)}(\Omega_{(2)}({\mathbb R}^D))$ for
$n\neq D$, $l\neq D$ and $l\neq 0$.
%
\subsection{Generalized Poincar\'e lemma in $\Omega^{(*,\ldots,*)}_{(*)}
({\mathbb R}^D)$}\label{section3}
%
Here we present the final result concerning our generalized Poincar\'e lemma:
\be
H^{(*,\ldots,*)}_{(*)}(\Omega_{(*)}({\mathbb R}^D))\cong 0\,,
\label{sowhat}\ee for diagrams obeying the assumption of the
Theorem ({\it cfr.} section \ref{GenPl}).
\\ \noindent 
Eqn (\ref{sowhat}) is really proved if one has the following
inductive progression:
\\ \noindent
{\it Under the assumption that 
\be
H^{(l_1,\ldots,l_{S-1},l_S)}_{(*)}(\Omega_{(S)}({\mathbb
R}^D))\cong 0 \ee $\forall (l_1,\ldots,l_{S-1})\in {\mathbb
Y}^{(S-1)}$ and $l_S$ fixed such
that $0<l_S<l_{S\!-\!1}$, \\ \noindent the following holds: 
\be
H^{(l_1,\ldots,l_{S-1},l_{S}+1)}_{(k)}(\Omega_{(S)}({\mathbb
R}^D))\simeq 0, \ee and \be
H^{(l_1,\ldots,l_{S-1},l_S,1)}_{(l)}(\Omega_{(S+1)}({\mathbb
R}^D))\simeq 0\,, \ee $0<l<S+2$.
}
\\ \\
 More explicitly, if the following statements are
satisfied: 
\ba &d^{I}\m(l_1,\ldots,l_{S-1},l_S)=0~~\forall
I\subset\{1,2,\ldots,S\}~\vert ~\# ~I=m&
\nonumber \\
&\Longrightarrow\m(l_1,\ldots,l_{S-1},l_S)=\sum_{J} d^{J}\n_J&
~~
\nonumber \\
&\forall J\subset\{1,2,\ldots,S\}~\vert~\#~
J=S\!+\!1\!-\!m~~{\rm{and}}~~
d^{J}\n_J\in\O_{(S)}^{(l_1,\ldots,l_{S-1},l_S)}({\mathbb
R}^D)\,,& \nn\ea it can be showed that these statements are also
true: \ba &~\bullet
d^{I}\m(l_1,\ldots,l_{S-1},l_S+1)=0~~\forall
I\subset\{1,2,\ldots,S\}~\vert~\# ~I=m&
\nonumber \\
&\Longrightarrow\m(l_1,\ldots,l_{S-1},l_S+1)=\sum_{J}
d^{J}\n_J& ~~
\nonumber \\
&\forall J\subset\{1,2,\ldots,S\}~\vert~\#~
J=S\!+\!1\!-\!m~~{\rm{and}}~~
d^{J}\n_J\in\O_{(S)}^{(l_1,\ldots,l_{S-1},l_S+1)}({\mathbb
R}^D)& \nn\ea and \ba &~\bullet
d^{I}\m(l_1,\ldots,l_{S-1},l_S,1)=0~~\forall
I\subset\{1,2,\ldots,S,S+1\}~\vert~\# ~I=m&
\nonumber \\
&\Longrightarrow\m(l_1,\ldots,l_{S-1},l_S,1)=\sum_{J}
d^{J}\n_J& ~~
\nonumber \\
&\forall J\subset\{1,\ldots,S,S+1\}~\vert~\#~
J=S\!+\!2\!-\!m~~{\rm{and}}~~
d^{\{J\}}\n_J\in\O_{(S)}^{(l_1,\ldots,l_S,1)}({\mathbb R}^D)\,.&
\nn\ea
We just have to follow the same lines as in sections 
\ref{subA1}, \ref{subA2} and \ref{subA3}.

%

\end{document}